\documentclass{article}

\usepackage{amsmath,amsthm,amssymb,bm,graphicx,float,mathrsfs}
\usepackage[hyphens]{url}
\usepackage[colorlinks,citecolor=blue,urlcolor=blue]{hyperref}
\pdfoutput=1

\def\P{{\rm P}}        
\def\E{{\rm E}}        
\def\SD{{\rm SD}}      
\def\T{\textsf{T}}     

\theoremstyle{plain}
\newtheorem{theorem}{Theorem}
\newtheorem{lemma}[theorem]{Lemma}

\allowdisplaybreaks

\title{Testing for dice control at craps}
\author{Stewart N. Ethier\thanks{\,Department of Mathematics, University of Utah. Email: \href{mailto:ethier@math.utah.edu}{ethier@math.utah.edu}.}}
\date{}

\begin{document}
\maketitle

\begin{abstract}
Dice control involves ``setting'' the dice and then throwing them carefully, in the hope of influencing the outcomes and gaining an advantage at craps.  How does one test for this ability?   To specify the alternative hypothesis, we need a statistical model of dice control.  Two have been suggested in the gambling literature, namely the Smith--Scott model and the Wong--Shackleford model.  Both models are parameterized by $\theta\in[0,1]$, which measures the shooter's level of control.  We propose and compare four test statistics: (a) the sample proportion of 7s; (b) the sample proportion of pass-line wins; (c) the sample mean of hand-length observations; and (d) the likelihood ratio statistic for a hand-length sample.    We want to test $H_0:\theta = 0$ (no control) versus $H_1:\theta > 0$ (some control).  We also want to test $H_0:\theta\le\theta_0$ versus $H_1:\theta>\theta_0$, where $\theta_0$ is the ``break-even point.''  For the tests considered we estimate the power, either by normal approximation or by simulation.\medskip

\end{abstract}

\section{Introduction}\label{sec:intro}

The idea of dice control as a legal method of advantage play at casino craps is almost as old as the game itself.  See the classic text \textit{How to Control Fair Dice} (c.~1922).  One approach is to ``set'' the dice and then attempt to throw them on-axis, to prevent two opposite faces of each die from appearing face up when the dice come to rest.  Complicating this is the requirement that the dice must bounce against the foam-rubber cushion at the opposite end of the table, which is not flat but rather covered with pyramid-shaped deflectors.  Dice control is regarded by the casino industry much as ESP is regarded by the psychology profession --- as a myth.

Yet there are those who claim to be successful dice controllers.  Some write books (Kononenko, 1998; Sharpshooter, 2002; Scoblete, 2003, 2015;  Scoblete and Dominator, 2010; Dominator, 2024) and some offer seminars/workshops (\url{https://www.goldentouchcraps.com}; \url{https://www.dicecoach.com}).  How can we assess the reliability of such claims?

A well-publicized challenge was held in 2004 intended to demonstrate an ability to reduce the probability of a 7 (Wong, 2005, Chap.~8).  The shooters were Stanford Wong and Little Joe Green and they threw the dice 500 times in a casino setting, obtaining 74 7s.  Viewed as a test of the null hypothesis $H_0: p(7)=1/6$ versus the alternative hypothesis $H_1: p(7)<1/6$, this resulted in a $p$-value of 0.144, which falls short of statistical significance.

Scott and Smith (2019) attempted to show that a dice-throwing machine could exercise some control.  They adopted what we refer to below as the $AA$ set (both dice rotate about the axis through their 1 and 6 faces), and applied a chi-squared goodness-of-fit test to the resulting single-die frequencies.  They failed to achieve statistical significance.  This experiment may have been motivated by Diaconis, Holmes, and Montgomery (2007), who created a coin-tossing machine with perfect predictability.  But clearly, controlling a coin toss caught in one's hand is not the same thing as controlling a pair of dice thrown onto a craps table.

Our aim here is to study statistical tests based on four relevant  test statistics: (a) the sample proportion of 7s; (b) the sample proportion of pass-line wins; (c) the sample mean of hand-length observations; and (d) the likelihood ratio statistic for a hand-length sample.  To explain these terms (pass line, hand length), we must briefly review the game of craps.  

The basic bet at craps is the \textit{pass-line bet}.  The initial roll is called the \textit{come-out roll}.  The bet is won if 7 or 11 appears (a \textit{natural}) and lost if 2, 3, or 12 appears (a \textit{craps number}).  Any other number (4, 5, 6, 8, 9, or 10) becomes the \textit{point}.  Once a point has been established, subsequent rolls are called \textit{point rolls}.  If the point reappears before a 7 appears, the bet is won.  If a 7 appears before the point reappears, the bet is lost.  The latter event is called a \textit{seven-out}.  A winning pass-line bet pays 1 to 1.  Once a pass-line bet is resolved, the next roll is a new come-out roll, and the process begins again.  Under the assumption of fair dice and no control, the expected gain from a one-unit pass-line bet is $-7/495$.

The \textit{shooter} is the player who rolls the dice. The shooter continues to roll until he or she sevens out, at which time the role of shooter is offered to the next player in clockwise order. The sequence of rolls by the shooter, from the initial come-out roll to the seven-out, is called the \textit{shooter’s hand}.  The \textit{length} of the shooter's hand is the number of rolls, inclusive of the initial come-out roll and the seven-out.  That number is a random variable assuming values in $\{2,3,4,\ldots\}$.  Under the assumption of fair dice and no control, its mean is $1671/196\approx8.52551$ and its variance is $1768701/38416\approx46.0407$.  

To properly formulate a test based on any of the aforementioned statistics, we need a statistical model of dice control.  Two such models have been suggested in the gambling literature, the Smith--Scott model (2018), which assumes an ability to roll the dice on-axis, and the Wong--Shackleford model (2005, 2023), which assumes an ability to roll the dice in a correlated way without favoring on-axis rolls.  Both models are parameterized by $\theta\in[0,1]$, which measures the shooter's level of control.

A third statistical model was proposed by Grosjean (2009, p.~471).  Here the probability of a 7 is $\theta$ and the remaining probability mass accrues to the other 10 dice totals proportionally;  more precisely,
\begin{equation*}
p(7,\theta)=\theta;\quad p(x,\theta)=(1-\theta)\frac{6-|x-7|}{30}, \quad x\in\{2,3,4,5,6,8,9,10,11,12\}.
\end{equation*}
Grosjean acknowledged that this model ``may not be a proper representation of the physical mechanism for altering'' the probability of a 7, and indeed, because it does not take dice setting explicitly into account, we regard it as lacking a plausible basis.

The interpretation of $\theta$ differs in our two models, so it may be useful to reparameterize the models with a parameter $\eta$ that has the same interpretation in both, such as letting $\eta$ be the expected gain from a one-unit pass-line bet.  In that case, we want to test $H_0:\eta=-7/495$ (no control) versus $H_1:\eta>-7/495$ (some control).  Alternatively, we may want to statistically demonstrate not just an ability to exercise some control but an ability to exercise control sufficient to provide an advantage at craps.  In that case we require the composite $H_0:\eta\le0$ (control not sufficient to provide an advantage) versus $H_1:\eta>0$ (control sufficient to provide an advantage).  A similar distinction was made in the context of testing for roulette-wheel bias (Ethier, 1982).

We introduce the Smith--Scott model in Section~\ref{sec:Smith-Scott}, arguing that its original formulation can be improved with a simple modification.  Following a brief discussion of dice setting, we introduce the Wong--Shackleford model in Section~\ref{sec:Wong-Shackleford}.  In Section~\ref{sec:reparam} we reparameterize both models so that the new parameter $\eta$ has the same interpretation in both; there are at least two ways to do this.  Section~\ref{sec:sample-proportions} discusses tests based on sample proportions.  In Section~\ref{sec:sample-mean} we consider a test based on the sample mean of $n$ hand-length observations, under both models, and we evaluate its power using a normal approximation.  In Section~\ref{sec:distrib-L} we derive the distribution of the length of the shooter's hand, under both models.  This is a generalization of a formula obtained by Ethier and Hoppe (2010) under the assumption of fair dice and no control.  In Section~\ref{sec:LRT} we study the likelihood ratio test and use simulation to estimate its power, under both models.  In Section~\ref{conclusions}, we summarize our conclusions.

We want to emphasize that this paper is concerned solely with methodology.  We have no actual data and take no position on the efficacy of dice control.

\section{The Smith--Scott model}\label{sec:Smith-Scott}

We begin by describing the dice-control model of Smith and Scott (2018).  It assumes a parameter $\theta\in[0,1]$ representing the shooter's level of control ($\theta=0$ means no control, $\theta=1$ means perfect control).  Each die can be ``set'' with the aim of reducing the probability of two specified opposite faces appearing.  An $A$ set attempts to reduce the chance of a 1 or a 6; a $B$ set attempts to reduce the chance of a 2 or a 5; and a $C$ set attempts to reduce the chance of a 3 or a 4.  With $S:=\{1,2,3,4,5,6\}$,
\begin{equation*}
A:=\{2,3,4,5\}, \quad
B:=\{1,3,4,6\}, \quad
C:=\{1,2,5,6\},
\end{equation*}
the distributions of single-die outcomes under these three die sets are
\begin{align*}
q_A(x,\theta)&:=(1-\theta)\text{UNIF}[S](x)+\theta\,\text{UNIF}[A](x),\\
q_B(x,\theta)&:=(1-\theta)\text{UNIF}[S](x)+\theta\,\text{UNIF}[B](x),\\
q_C(x,\theta)&:=(1-\theta)\text{UNIF}[S](x)+\theta\,\text{UNIF}[C](x),
\end{align*}
for $x\in\{1,2,3,4,5,6\}$.
In other words, the distributions $q_A$, $q_B$, and $q_C$ are $(1-\theta,\theta)$ mixtures of the results of the roll of a standard die and that of a die that rolls on a horizontal axis through two opposite faces, preventing those two faces from appearing face up.

The controlled dice shooter sets each of the two dice and therefore has six possible dice sets, $AA$, $AB$, $AC$, $BB$, $BC$, and $CC$.  The distribution of dice totals is given by convolution, namely
\begin{align*}
q_{AA}(x,\theta)&:=[q_A(\,\cdot\,,\theta)*q_A(\,\cdot\,,\theta)](x),\\
q_{AB}(x,\theta)&:=[q_A(\,\cdot\,,\theta)*q_B(\,\cdot\,,\theta)](x),\\
q_{AC}(x,\theta)&:=[q_A(\,\cdot\,,\theta)*q_C(\,\cdot\,,\theta)](x),\\
q_{BB}(x,\theta)&:=[q_B(\,\cdot\,,\theta)*q_B(\,\cdot\,,\theta)](x),\\
q_{BC}(x,\theta)&:=[q_B(\,\cdot\,,\theta)*q_C(\,\cdot\,,\theta)](x),\\
q_{CC}(x,\theta)&:=[q_C(\,\cdot\,,\theta)*q_C(\,\cdot\,,\theta)](x),
\end{align*}
for $x\in\{2,3,4,\ldots,12\}$.  This is the original Smith--Scott model.  

We believe that this model is conceptually flawed by its implicit assumption that the two dice behave independently.  When the two dice are set and thrown together, it seems intuitively clear that independence must fail if $0<\theta<1$.

We propose replacing the convolution of mixtures by the mixture of convolutions.  In more detail, with no control, the distribution of dice totals is given by the convolution
\begin{equation*}
p_{SS}(x):=[\text{UNIF}[S](\cdot)*\text{UNIF}[S](\cdot)](x)=\frac{6-|x-7|}{36},\\
\end{equation*}
for $x\in\{2,3,4,\ldots,12\}$. With perfect control, it is given by the convolution
\begin{align*}
p_{AA}(x)&:=[\text{UNIF}[A](\cdot)*\text{UNIF}[A](\cdot)](x),\\
p_{AB}(x)&:=[\text{UNIF}[A](\cdot)*\text{UNIF}[B](\cdot)](x),\\
p_{AC}(x)&:=[\text{UNIF}[A](\cdot)*\text{UNIF}[C](\cdot)](x),\\
p_{BB}(x)&:=[\text{UNIF}[B](\cdot)*\text{UNIF}[B](\cdot)](x),\\
p_{BC}(x)&:=[\text{UNIF}[B](\cdot)*\text{UNIF}[C](\cdot)](x),\\
p_{CC}(x)&:=[\text{UNIF}[C](\cdot)*\text{UNIF}[C](\cdot)](x),
\end{align*}
for $x\in\{2,3,4,\ldots,12\}$.  These distributions were first evaluated by Kononenko (1998, Chap.\ 19).  Independently, they were subsequently evaluated by Smith and Scott (2018, Table 2).

Now we introduce the parameter $\theta\in[0,1]$ representing the shooter's level of control as before.  We define
\begin{align*}
p_{AA}(x,\theta)&:=(1-\theta)p_{SS}(x)+\theta\,p_{AA}(x),\\
p_{AB}(x,\theta)&:=(1-\theta)p_{SS}(x)+\theta\,p_{AB}(x),\\
p_{AC}(x,\theta)&:=(1-\theta)p_{SS}(x)+\theta\,p_{AC}(x),\\
p_{BB}(x,\theta)&:=(1-\theta)p_{SS}(x)+\theta\,p_{BB}(x),\\
p_{BC}(x,\theta)&:=(1-\theta)p_{SS}(x)+\theta\,p_{BC}(x),\\
p_{CC}(x,\theta)&:=(1-\theta)p_{SS}(x)+\theta\,p_{CC}(x),
\end{align*}
for $x\in\{2,3,4,\ldots,12\}$.  This is the modified Smith--Scott model, but hereafter we will omit the word ``modified'' and call it simply the \textit{Smith--Scott model}, believing that its original authors would endorse this modification.

An equivalent formulation involves first finding the joint distribution of the two dice.  Thus,
\begin{align*}
p_{AA}(x,y,\theta)&:=(1-\theta)\text{UNIF}[S\times S](x,y)+\theta\,\text{UNIF}[A\times A](x,y),\\
p_{AB}(x,y,\theta)&:=(1-\theta)\text{UNIF}[S\times S](x,y)+\theta\,\text{UNIF}[A\times B](x,y),\\
p_{AC}(x,y,\theta)&:=(1-\theta)\text{UNIF}[S\times S](x,y)+\theta\,\text{UNIF}[A\times C](x,y),\\
p_{BB}(x,y,\theta)&:=(1-\theta)\text{UNIF}[S\times S](x,y)+\theta\,\text{UNIF}[B\times B](x,y),\\
p_{BC}(x,y,\theta)&:=(1-\theta)\text{UNIF}[S\times S](x,y)+\theta\,\text{UNIF}[B\times C](x,y),\\
p_{CC}(x,y,\theta)&:=(1-\theta)\text{UNIF}[S\times S](x,y)+\theta\,\text{UNIF}[C\times C](x,y),
\end{align*}
for $x,y\in\{1,2,3,4,5,6\}$, followed by the convolution-like formulas
\begin{align*}
p_{AA}(z,\theta)&:=\sum_x p_{AA}(x,z-x,\theta),\\
p_{AB}(z,\theta)&:=\sum_x p_{AB}(x,z-x,\theta),\\
p_{AC}(z,\theta)&:=\sum_x p_{AC}(x,z-x,\theta),\\
p_{BB}(z,\theta)&:=\sum_x p_{BB}(x,z-x,\theta),\\
p_{BC}(z,\theta)&:=\sum_x p_{BC}(x,z-x,\theta),\\
p_{CC}(z,\theta)&:=\sum_x p_{CC}(x,z-x,\theta),
\end{align*}
for $z\in\{2,3,4,\ldots,12\}$.  Note that we can interpret $\theta$ to be the proportion by which off-axis rolls are reduced by a skilled shooter.

Under the model, the two dice do not behave independently.  For example, let $X$ and $Y$ be the results of the two dice.  Then, under the $AA$ set,
\begin{equation*}
\P_\theta(X\in A,\,Y\in A)=\sum_{x\in A}\sum_{y\in A}p_{AA}(x,y,\theta)=(1-\theta)\frac{16}{36}+\theta\frac{16}{16}=\frac{4+5\,\theta}{9},
\end{equation*}
while
\begin{align*}
\P_\theta(X\in A)\,\P_\theta(Y\in A)&=\sum_{x\in A}\sum_{y\in S}p_{AA}(x,y,\theta)\sum_{x\in S}\sum_{y\in A}p_{AA}(x,y,\theta) \\
&=\bigg((1-\theta)\frac{24}{36}+\theta\frac{16}{16}\bigg)^2=\bigg(\frac{2+\theta}{3}\bigg)^2=\frac{4+4\,\theta+\theta^2}{9},
\end{align*}
whereby
\begin{equation*}
\P_\theta(X\in A,\,Y\in A)-\P_\theta(X\in A)\,\P_\theta(Y\in A)=\frac{\theta(1-\theta)}{9}.
\end{equation*}

Now let us consider which of these dice sets is optimal at craps.

When the point is 4 or 10, the probability of making the point is uniquely maximized by the dice set $AC$, with the probability increasing linearly from 1/3 ($\theta=0$) to 1/2 ($\theta=1$).

When the point is 5 or 9, the probability of making the point is maximized by each of the dice sets $AB$, $AC$, and $BC$, with the probability increasing nonlinearly from 2/5 ($\theta=0$) to 1/2 ($\theta=1$).

When the point is 6 or 8, the probability of making the point is uniquely maximized by the dice set $AB$, with the probability increasing nonlinearly from 5/11 ($\theta=0$) to 3/5 ($\theta=1$).

On the come-out roll, the probability of winning a pass-line bet (assuming an optimal set after each point) is uniquely maximized by the dice set $AA$, with the probability increasing nonlinearly from 244/495 ($\theta=0$) to 53/80 ($\theta=1$).

Under the S--S model, an optimal strategy is therefore as follows: the $AA$ set is used on the come-out roll, the $AC$ set is used if the point is 4, 5, 9, or 10, and the $AB$ set is used if the point is 6 or 8.  We consider this strategy to be part of the S--S model.  (The same strategy is optimal for the original S--S model, before our modification, at least for $\theta>0.272898$.  For $\theta<0.272898$, the $AA$ set is preferable when the point is 6 or 8.  Of course, the choice of dice set cannot depend on the unknown $\theta$.)

Let $G$ be the gain from a one-unit pass-line bet. Under the S--S model, its expectation is
\begin{align}\label{EG-SS}
\E_\theta[G]&=2\bigg[p_{AA}(7,\theta)+p_{AA}(11,\theta)+\sum_{x=4,5,9,10}p_{AA}(x,\theta)\frac{p_{AC}(x,\theta)}{p_{AC}(x,\theta)+p_{AC}(7,\theta)}\nonumber\\
&\qquad\quad{}+\sum_{x=6,8}p_{AA}(x,\theta)\frac{p_{AB}(x,\theta)}{p_{AB}(x,\theta)+p_{AB}(7,\theta)}\bigg]-1\nonumber\\
&=-\frac{448 - 10064\,\theta + 72\,\theta^2 + 68\,\theta^3 - \theta^4}{72 (10 - \theta) (44 + \theta)},
\end{align}
which equals 0 when $\theta$ is a root of the quartic polynomial in the numerator.  Hence the break-even point is $\theta_0\approx0.0445299$.  (Under the original S--S model, before our modification, the break-even point was $\theta_0\approx0.0803061$ or $\theta_0\approx0.0948562$, depending on whether the $AA$ or $AB$ set is used when the point is 6 or 8.)

\section{The Wong--Shackleford model}\label{sec:Wong-Shackleford}

\begin{figure}[H]
\begin{center}
\includegraphics[width=2.2in]{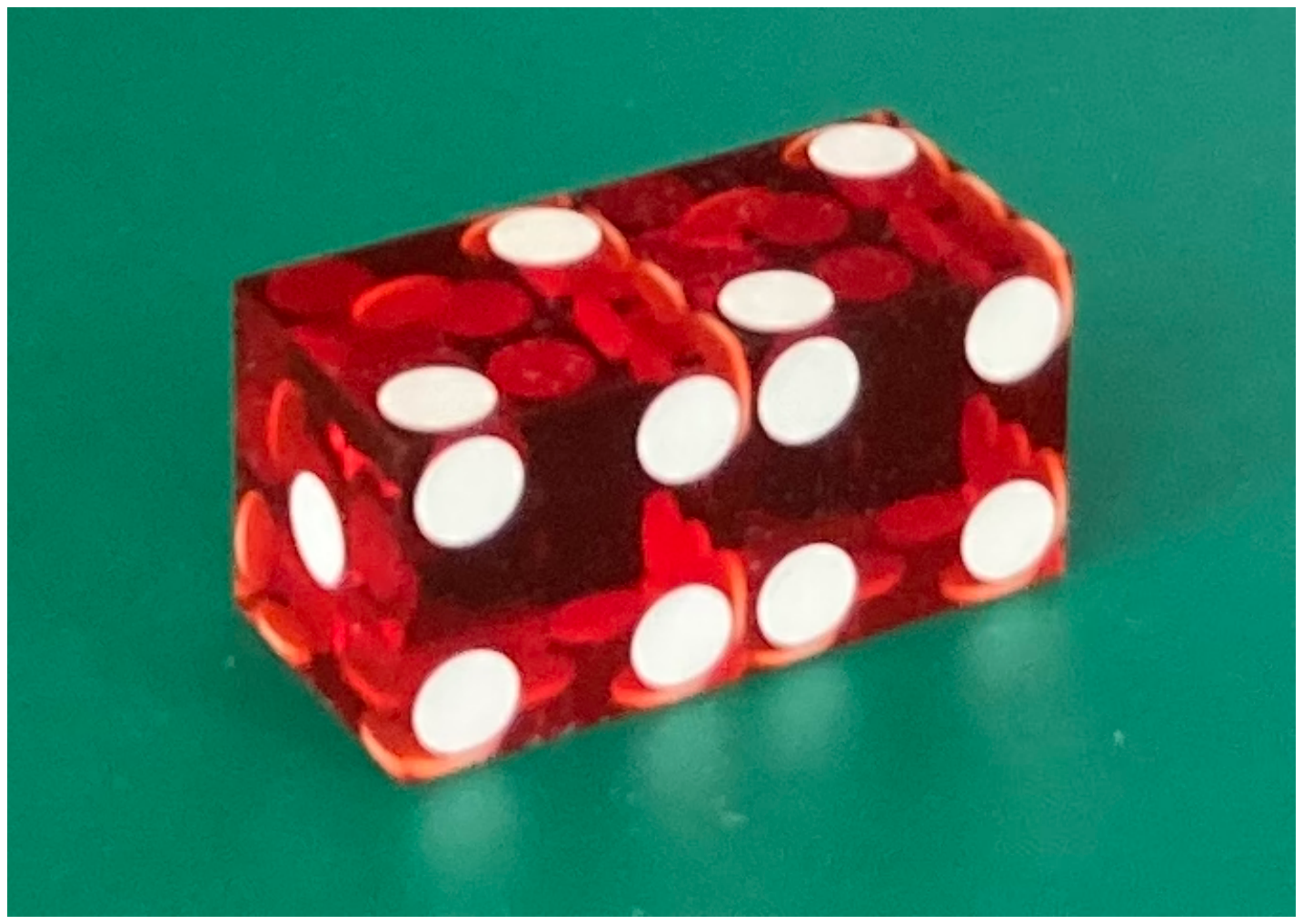}\qquad\includegraphics[width=2.2in]{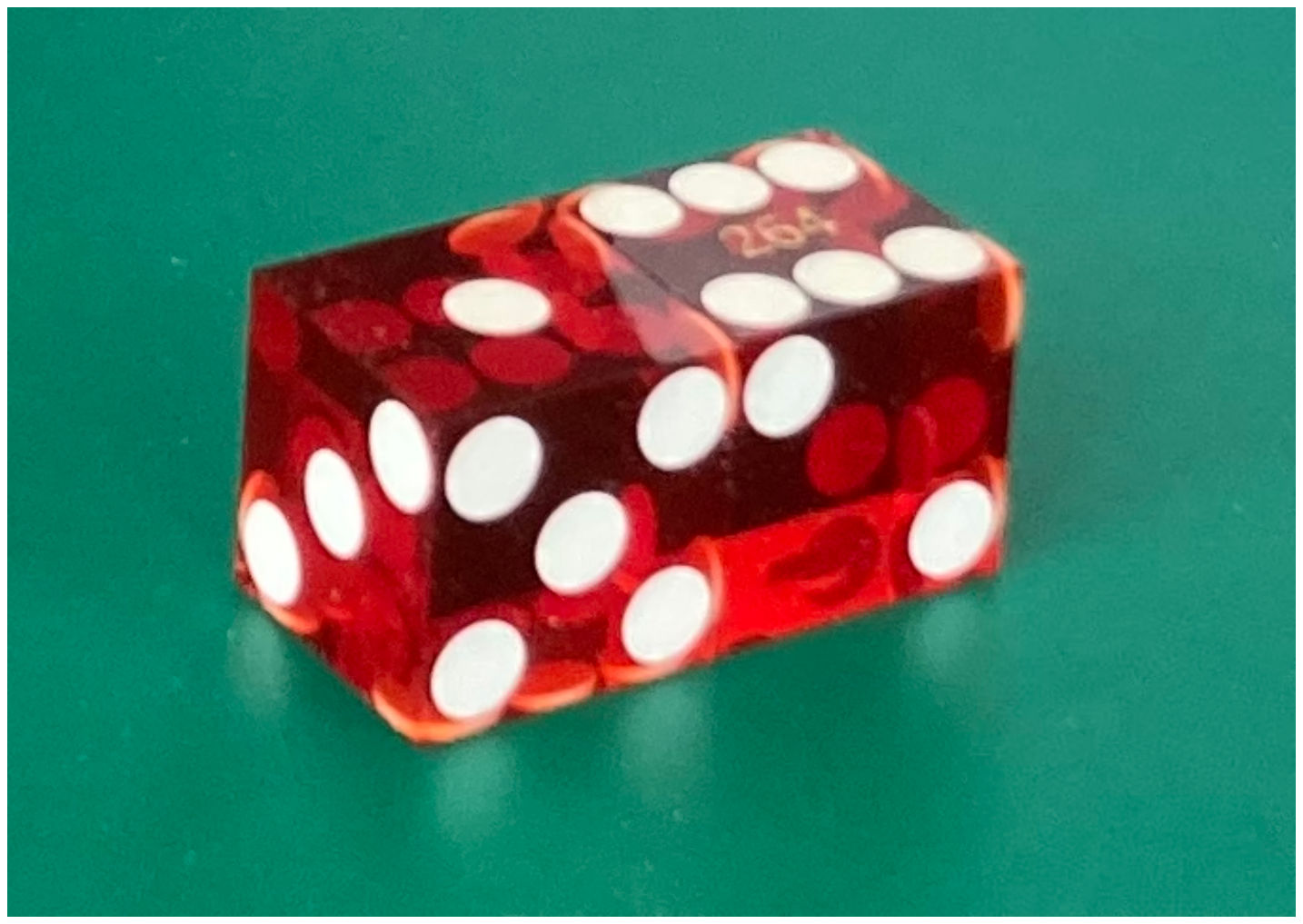}
\end{center}
\caption{\label{2424-1562}The Hard Ways Set \#1 or $(2,4,2,4)$, and the Sevens Set \#1 or $(1,5,6,2)$.}
\end{figure}

First, let us briefly discuss the traditional approach to dice setting.  (The method of Section~\ref{sec:Smith-Scott} is somewhat unconventional, in that only the axis about which the dice rotate was specified in setting the dice.)  A dice set (or dice arrangement) can be described by four faces of the dice, denoted by $(a,b,c,d)$, with $a$ and $b$ being the top and front faces of the left die, and $c$ and $d$ being the top and front faces of the right die.  Keep in mind that opposite faces sum to 7, so we write
\begin{equation*}
\bar a:=7-a
\end{equation*}
for the face opposite face $a$.
For example, the Hard Ways Set \#1 of Figure~\ref{2424-1562} below is $(2,4,2,4)$, and the Sevens Set \#1 of the same figure is $(1,5,6,2)$.  Both belong to the set of all dice sets
\begin{equation*}
D:=\{(a,b,c,d): a,b,c,d\in\{1,2,3,4,5,6\},\, b\notin \{a,\bar a\},\, d\notin \{c,\bar c\}\}.
\end{equation*}

But there are three permutations of $D$ that map dice sets to equivalent dice sets.  First, there is rotation about the axis,
\begin{equation*}
\sigma_1(a,b,c,d)=(b,\bar a,d,\bar c);
\end{equation*}
second, there is axis reversal,
\begin{equation*}
\sigma_2(a,b,c,d)=(c,\bar d,a,\bar b);
\end{equation*}
and third, there is interchange of dice,
\begin{equation*}
\sigma_3(a,b,c,d)=(c,d,a,b).
\end{equation*}
$\sigma_1$ is of order 4, while $\sigma_2$ and $\sigma_3$ are of order 2.  The group $G$ of order 16 generated by $\sigma_1$, $\sigma_2$, and $\sigma_3$ defines a group action on $D$ whose orbits partition $D$ into equivalence classes.  As Wong (2005, p.~154) showed, there are 45 equivalence classes, 18 of size 8 and 27 of size 16, accounting for all $6\cdot4\cdot6\cdot4=576$ elements of $D$.  The Hard Ways Set \#1 and the Sevens Set \#1 appear in Wong's list of the 45 distinct dice sets as $(2,3,2,3)$ and $(1,2,6,5)$ because, to represent a given equivalence class, Wong chose its first element in lexicographic order.  The number of equivalence classes, 45, can also be surmised from Burnside's lemma:
\begin{equation*}
|D/G|=\frac{1}{|G|}\sum_{\sigma\in G}|D^\sigma|=\frac{(1)576+(6)24+(9)0}{16}=\frac{720}{16}=45.
\end{equation*}
(Here $|D/G|$ is the number of orbits and $D^\sigma:=\{(a,b,c,d)\in D:\sigma(a,b,c,d)=(a,b,c,d)\}$.)

To connect these results with those of the preceding section, we observe that rotation of just the left die, as in
\begin{equation*}
\sigma_4(a,b,c,d):=(b,\bar a,c,d),
\end{equation*}
and reversal of just the left die, as in
\begin{equation*}
\sigma_5(a,b,c,d):=(a,\bar b,c,d),
\end{equation*}
result in an equivalent dice set in the sense of the preceding section.  The 45 equivalence classes identified above become just six.  $AA$, $BB$, and $CC$ are each obtained by combining six equivalence classes of size 8 and one of size 16 (64 elements total), and $AB$, $AC$, and $BC$ are each obtained by combining eight equivalence classes of size 16 (128 elements total).  And Burnside's lemma gives
\begin{equation*}
|D/G'|=\frac{1}{|G'|}\sum_{\sigma\in G'}|D^\sigma|=\frac{(1)576+(8)24+(119)0}{128}=\frac{768}{128}=6.
\end{equation*}

The original formulation of a statistical model by Wong (2005, Chap.~6) was slightly ambiguous but clarified in the reinterpretation by Shackleford (2023).  Unlike in the S--S model, on-axis tosses are not more likely than for random shooters, but when a toss is on-axis, the respective rotations of the two dice about their axes are correlated.  If these rotations differ by 0 degrees, the toss is called a \textit{zero pitch}.  If they differ by 90 degrees in either direction, the toss is called a \textit{single pitch}.  If they differ by 180 degrees, the toss is called a \textit{double pitch}.  Wong's hypothesis was that the chance of a double pitch is reduced by a skilled shooter.  Shackleford's clarification is that the reduced chance of a double pitch results in an augmented chance of the corresponding zero pitch, with the chance of a single pitch left unaffected.

If the set $(a,b,c,d)$ is adopted, the joint distribution of the dice is
\begin{align*}
p_{abcd}(x,y,\theta)&:=\frac{1}{36}+\bm1_{\{(a,c),(b,d),(\bar a,\bar c),(\bar b,\bar d)\}}(x,y)\frac{\theta}{36}\\
\noalign{\smallskip}
&\qquad\quad{}-\bm1_{\{(a,\bar c),(b,\bar d),(\bar a,c),(\bar b,d)\}}(x,y)\frac{\theta}{36}
\end{align*} 
for $x,y\in\{1,2,3,4,5,6\}$, the first indicator function corresponding to the zero pitches and the second corresponding to the double pitches.  The parameter $\theta$, which is the proportion of double pitches converted to zero pitches, is what Shackleford called the \textit{skill factor}.

It follows that the distribution of dice totals is
\begin{align*}
p_{abcd}(x,\theta)&=\frac{6-|x-7|}{36}\\
&\quad{}+\big(\bm1_{\{a+c\}}(x)+\bm1_{\{b+d\}}(x)+\bm1_{\{\bar a+\bar c\}}(x)+\bm1_{\{\bar b+\bar d\}}(x)\big)\frac{\theta}{36}\\
&\quad{}-\big(\bm1_{\{a+\bar c\}}(x)+\bm1_{\{b+\bar d\}}(x)+\bm1_{\{\bar a+c\}}(x)+\bm1_{\{\bar b+d\}}(x)\big)\frac{\theta}{36}
\end{align*}
for $x\in\{2,3,4,\ldots,12\}$.

Shackleford (2023) claims to have shown that, under this model, it is optimal to use the Sevens Set \#1 or $(1,5,6,2)$ on the come-out roll and the Hard Ways Set \#1 or $(2,4,2,4)$ on point rolls, regardless of the point.  We have confirmed his result and consider this strategy to be part of the W--S model.

Let $G$ be the gain from a one-unit pass-line bet. Under the W--S model, its expectation is 
\begin{align}\label{EG-WS}
\E_\theta[G]
&=2\bigg[p_{1562}(7,\theta)+p_{1562}(11,\theta) \nonumber\\
\noalign{\vglue-2mm}
&\qquad\quad{}+\sum_{x=4,5,6,8,9,10}p_{1562}(x,\theta)\frac{p_{2424}(x,\theta)}{p_{2424}(x,\theta)+p_{2424}(7,\theta)}\bigg]-1\nonumber\\
&=-\frac{21 - 682\,\theta + 361\,\theta^2 - 42\,\theta^3}{27 (5 - 2\,\theta) (11 - 3\,\theta)},
\end{align}
which equals 0 when $\theta$ is a root of the cubic polynomial in the numerator. Hence the break-even point is $\theta_0\approx0.0313088$.

\section{Reparameterization of the models}\label{sec:reparam}

Later we will want to compare the power of certain tests under the two models.  But for such a comparison to be useful, the parameter $\theta$ should have the same interpretation in both models.  In the S--S model, $\theta$ is the proportion by which off-axis rolls are reduced, and in the W--S model, $\theta$ is the proportion of double-pitch on-axis rolls converted to zero-pitch on-axis rolls.  These two interpretations of $\theta$ are not the same.  Thus, we propose reparameterizing both models in such a way that the new parameter $\eta$ has the same interpretation in each.

There are at least two ways to do this.  The first method is simple but not entirely satisfactory.  The second method seems more natural but is also much more complicated.

The first method is to parameterize both models by the reciprocal of the probability of rolling a 7 when a point has been established.  In the S--S model, the new parameter is
\begin{equation*}
\eta=\rho(\theta):=\frac{1}{p_{AB}(7,\theta)}=\frac{1}{p_{AC}(7,\theta)}=\frac{24}{4-\theta}\in[6,8].
\end{equation*}
In the W--S model, it is
\begin{equation*}
\eta=\rho(\theta):=\frac{1}{p_{2424}(7,\theta)}=\frac{18}{3-2\theta}\in[6,18].
\end{equation*}
For $\eta\in[6,8]$ (with $[6,8]$ being the intersection of the two $\rho$ ranges), we replace $\theta$ in the S--S model by
\begin{equation*}
\theta=\rho^{-1}(\eta)=4-\frac{24}{\eta}\in[0,1]
\end{equation*}
and $\theta$ in the W--S model by
\begin{equation*}
\theta=\rho^{-1}(\eta)=\frac32-\frac{9}{\eta}\in\bigg[0,\frac38\bigg].
\end{equation*}

The only problem with this reparameterization is that the quantity that is equalized between the two models, namely the probability of rolling a 7 on a point roll, is not the most important quantity because it ignores come-out rolls.  A more important one is the probability of winning a pass-line bet or, equivalently, the expected gain from a one-unit pass-line bet.

In the reparameterization, the two models are still quite different.  For example, the break-even point in the S--S model is $\eta_0=\rho(\theta_0)=24/(4-\theta_0)\approx6.06755$, and in the W--S model it is $\eta_0=\rho(\theta_0)=18/(3-2\theta_0)\approx6.12790$.

A second method is to parameterize both models by the expected gain from a one-unit pass-line bet.  In the S--S model this is, by \eqref{EG-SS},
\begin{equation*}
\eta=\rho(\theta)=\E_\theta[G]=-\frac{448 - 10064\,\theta + 72\,\theta^2 + 68\,\theta^3 - \theta^4}{72 (10 - \theta) (44 + \theta)}\in\bigg[-\frac{7}{495},\frac{13}{40}\bigg],
\end{equation*}
and in the W--S model it is, by \eqref{EG-WS},
\begin{equation*}
\eta=\rho(\theta)=\E_\theta[G]=-\frac{21 - 682\,\theta + 361\,\theta^2 - 42\,\theta^3}{27 (5 - 2\,\theta) (11 - 3\,\theta)}\in\bigg[-\frac{7}{495},\frac{19}{36}\bigg].
\end{equation*}
We can solve these two equations for $\theta$ in terms of $\eta\in[-7/495,13/40]$ (with $[-7/495,13/40]$ being the intersection of the two $\rho$ ranges) using the quartic and cubic formulas, or using \textit{Mathematica}.  Unfortunately, the resulting formulas for $\theta=\rho^{-1}(\eta)$ are complicated and inelegant, so we do not attempt to reproduce them here, but we will use them in what follows.

\section{Tests based on sample proportions}\label{sec:sample-proportions}

Suppose we want to test the ability to reduce the chance of rolling a 7 on a point roll.  The first reparameterization is then relevant.  We test $H_0:\eta=6$ versus $H_1:\eta>6$ using the test statistic $\hat p$, the sample proportion of 7s.  Under $H_0$,
\begin{equation*}
\frac{\hat p-1/6}{\sqrt{(1/6)(1-1/6)/n}}
\end{equation*}
is approximately standard normal, so a test of approximate size $\alpha$ has critical region equal to
\begin{equation*}
\hat p<1/6-\Phi^{-1}(1-\alpha)\sqrt{(1/6)(1-1/6)/n},
\end{equation*}
where $\Phi$ is the cumulative distribution function of the standard normal.  Its power at $\eta$ is 
\begin{align*}
\beta(\eta)&=\P_\eta\big(\hat p<1/6-\Phi^{-1}(1-\alpha)\sqrt{(1/6)(1-1/6)/n}\,\big) \\
&\approx\Phi\bigg(\frac{\sqrt{n}(1/6-1/\eta)-\Phi^{-1}(1-\alpha)\sqrt{(1/6)(1-1/6)}}{\sqrt{(1/\eta)(1-1/\eta)}}\bigg).
\end{align*}
Notice that this applies to both of our models.  

We could also test $H_0:\eta\le\eta_0$ versus $H_1:\eta>\eta_0$, and here the power at $\eta$ would be 
\begin{align*}
\beta(\eta)&=\P_\eta\big(\hat p<1/\eta_0-\Phi^{-1}(1-\alpha)\sqrt{(1/\eta_0)(1-1/\eta_0)/n}\,\big) \\
&\approx\Phi\bigg(\frac{\sqrt{n}(1/\eta_0-1/\eta)-\Phi^{-1}(1-\alpha)\sqrt{(1/\eta_0)(1-1/\eta_0)}}{\sqrt{(1/\eta)(1-1/\eta)}}\bigg),
\end{align*}
but the value of the break-even point $\eta_0$ differs in the two models (see the preceding section).

We might also want to test the shooter's ability to win at craps.  Here the second parameterization is preferable.  With $\eta$ denoting the expected gain from a one-unit pass-line bet, we can test $H_0:\eta=-7/495$ versus $H_1:\eta>-7/495$. Let $\hat p$ be the proportion of pass-line wins in $n$ pass-line decisions.  Then, under $H_0$,
\begin{equation*}
\frac{2\hat p-1-(-7/497)}{\sqrt{(1-(-7/495)^2)/n}}
\end{equation*}
is approximately standard normal.  Thus, the critical region for a test of significance level approximately $\alpha$ is
\begin{equation*}
2\hat p-1>-7/495+\Phi^{-1}(1-\alpha)\sqrt{(1-(-7/495)^2)/n}
\end{equation*}
and its power at the alternative $\eta$ is
\begin{align*}
\beta(\eta)&=\P_\eta\big(2\hat p-1>-7/495+\Phi^{-1}(1-\alpha)\sqrt{(1-(-7/495)^2)/n}\,\big) \\
&\approx1-\Phi\bigg(\frac{\sqrt{n}(-7/495-\eta)+\Phi^{-1}(1-\alpha)\sqrt{1-(-7/495)^2}}{\sqrt{1-\eta^2}}\bigg).
\end{align*}
This applies to both models.

For the case of the composite null hypothesis $H_0:\eta\le0$ versus the alternative hypothesis $H_1:\eta>0$, the power at $\eta$ becomes
\begin{align*}
\beta(\eta)&=\P_\eta\big(2\hat p-1>\Phi^{-1}(1-\alpha)/\sqrt{n}\,\big) \\
&\approx1-\Phi\bigg(\frac{\sqrt{n}(-\eta)+\Phi^{-1}(1-\alpha)}{\sqrt{1-\eta^2}}\bigg).
\end{align*}
Again, this applies to both models.

Other possible test statistics include those specific to one model or the other.  For example, under the S--S model, the sample proportion of on-axis throws would be relevant, but we do not pursue this.

\section{Tests based on the sample mean $\overline{L}$}\label{sec:sample-mean}

To see whether a test based on hand-length observations makes sense, we evaluate the mean and variance of the length $L$ of the shooter's hand, first under the S--S model.   These are just the mean and variance of the absorption time of a five-state Markov chain.  Its state space is $\{\text{co},\text{p4-10},\text{p5-9},\text{p6-8},\text{7o}\}$, where the states are interpreted as: shooter is coming out, point 4 or 10 is established, point 5 or 9 is established, point 6 or 8 is established, and shooter has sevened out.  (We are implicitly using the fact that, regardless of $\theta$, the totals 4 and 10 have equal probabilities; the same is true of 5 and 9 and of 6 and 8.)  Its one-step transition matrix is
\begin{equation}\label{SS-P_theta}
\bm P_\theta:=(1-\theta)\frac{1}{36}
\begin{pmatrix}
 12 & 6 & 8 & 10 & 0 \\
3 & 27 & 0 & 0 & 6 \\
4 & 0 & 26 & 0 & 6 \\
5 & 0 & 0 & 25 & 6 \\
0 & 0 & 0 & 0 & 36
\end{pmatrix}
+\theta\frac{1}{16}
\begin{pmatrix}
4 & 2 & 4 & 6 & 0 \\
2 & 12 & 0 & 0 & 2 \\
2 & 0 & 12 & 0 & 2 \\
3 & 0 & 0 & 11 & 2 \\
0 & 0 & 0 & 0 & 16
\end{pmatrix},\quad
\end{equation} 
and its initial state is co.   Here row co is determined by $p_{AA}(\,\cdot\,,\theta)$, rows p4-10 and p5-9 are determined by $p_{AC}(\,\cdot\,,\theta)$, row p6-8 is determined by $p_{AB}(\,\cdot\,,\theta)$, and row 7o is determined by the assumption that state 7o is absorbing.

Let $\bm Q_\theta$ denote the principal $4\times4$ submatrix of $\bm P_\theta$ corresponding to the four transient states.  Then, with $\bm M_\theta:=(\bm I_4-\bm Q_\theta)^{-1}$ denoting the \textit{fundamental matrix}, the mean of $L$ is
\begin{equation*}
\E_\theta[L]=(\bm M_\theta\bm1)_{\text{co}}=\frac{24 (8912 + 132\,\theta - 54\,\theta^2 +\,\theta^3)}{(4 - \theta) (8 + \theta)(28 - \theta)^2},
\end{equation*}
where $\bm1:=(1,1,1,1)^\T$ (Kemeny and Snell, 1976, Theorem 3.3.5).  This function equals 
\begin{equation*}
\frac{1671}{196}\approx8.52551\text{ at }\theta=0\;\;\text{and}\;\;\frac{296}{27}\approx10.9630\text{ at }\theta=1,
\end{equation*}
and it is increasing on $[0,1]$.

The variance of $L$ can also be computed in a similar way.  We find that (Kemeny and Snell, 1976, Theorem 3.3.5)
\begin{align*}
\text{Var}_\theta(L)&= [(2\bm M-\bm I_4)\bm M\bm1]_{\text{co}}-[(\bm M\bm1)_{\text{co}}]^2\nonumber\\
&=24 (44 + \theta) (27441664 + 4461696\,\theta - 441024\,\theta^2 - 34064\,\theta^3 \nonumber\\
&\qquad{}+ 5652\,\theta^4 - 174\,\theta^5 + \theta^6)/[(4 - \theta)^2 (8 + \theta)^2 (28 - \theta)^4].
\end{align*}
This function equals 
\begin{equation*}
\frac{1768701}{38416}\approx46.0407\text{ at }\theta=0\;\text{and}\;\frac{63880}{729}\approx87.6269\text{ at }\theta=1,
\end{equation*}
and it is increasing on $[0,1]$.  

We can similarly evaluate the mean and variance of $L$ under the W--S model.  The one-step transition matrix is
\begin{equation}\label{WS-P_theta}
\bm P_\theta:=(1-\theta)\frac{1}{36}\begin{pmatrix}
 12 & 6 & 8 & 10 & 0 \\
3 & 27 & 0 & 0 & 6 \\
4 & 0 & 26 & 0 & 6 \\
5 & 0 & 0 & 25 & 6 \\
0 & 0 & 0 & 0 & 36
\end{pmatrix}
+\theta\frac{1}{36}
\begin{pmatrix}
14 & 4 & 8 & 10 & 0 \\
4 & 30 & 0 & 0 & 2 \\
4 & 0 & 30 & 0 & 2 \\
6 & 0 & 0 & 28 & 2 \\
0 & 0 & 0 & 0 & 36 
\end{pmatrix}.\quad
\end{equation} 
Here row co is determined by $p_{1562}(\,\cdot\,,\theta)$, rows p4-10, p5-9, and p6-8 are determined by $p_{2424}(\,\cdot\,,\theta)$, and row 7o is determined by the assumption that state 7o is absorbing.

For the mean of $L$ we obtain
\begin{equation*}
\E_\theta[L]=\frac{9 (557 - 281\,\theta + 30\,\theta^2)}{(3 - 2\,\theta) (196 - 85\,\theta + 6\,\theta^2)}.
\end{equation*}
This function equals 
\begin{equation*}
\frac{1671}{196}\approx8.52551\text{ at }\theta=0\;\;\text{and}\;\;\frac{306}{13}\approx23.5385\text{ at }\theta=1,
\end{equation*}
and it is increasing on $[0,1]$.

For the variance of $L$ we obtain
\begin{align*}
\text{Var}_\theta(L)&=9 (5306103 - 5622318\,\theta + 1933097\,\theta^2 - 153376\,\theta^3 \nonumber\\
&\qquad\qquad{}- 39214\,\theta^4 + 7176\,\theta^5 - 360\,\theta^6)\nonumber\\ 
&\qquad{}/[(3 - \theta) (3 - 2\,\theta)^2 (196 - 85\,\theta + 6\,\theta^2)^2].
\end{align*}
This function equals 
\begin{equation*}
\frac{1768701}{38416}\approx46.0407\text{ at }\theta=0\;\text{and}\;\frac{79506}{169}\approx470.450\text{ at }\theta=1,
\end{equation*}
and it is increasing on $[0,1]$.

A test of the null hypothesis that $\theta=0$ could be based on the sample mean $\overline{L}$.   Let $\alpha$ be the desired significance level.  Then 
\begin{equation*}
\frac{\overline{L}-\E_0[L]}{\text{SD}_0(L)/\sqrt{n}}
\end{equation*}
is approximately standard normal under $H_0$, so we reject $H_0$ if
\begin{equation*}
\overline{L}>\E_0[L]+\Phi^{-1}(1-\alpha)\frac{\text{SD}_0(L)}{\sqrt{n}}.
\end{equation*}
The power of the test under the alternative $\theta$ is then
\begin{align*}
\beta(\theta)&=\P_\theta(\overline{L}>\E_0[L]+\Phi^{-1}(1-\alpha)\SD_0(L)/\sqrt{n}) \\
&=\P_\theta\bigg(\frac{\overline{L}-\E_\theta[L]}{\SD_\theta(L)/\sqrt{n}}>\frac{\E_0[L]-\E_\theta[L]+\Phi^{-1}(1-\alpha)\text{SD}_0(L)/\sqrt{n}}{\text{SD}_\theta(L)/\sqrt{n}}\bigg) \\
&\approx1-\Phi\bigg(\frac{\sqrt{n}(\E_0[L]-\E_\theta[L])+\Phi^{-1}(1-\alpha)\text{SD}_0(L)}{\text{SD}_\theta(L)}\bigg).
\end{align*}

In the case of a composite null hypothesis, we will need the following lemma.

\begin{lemma}\label{Lemma1}
Under either the S--S model or the W--S model, $\P_\theta(\overline{L}>x)$ is nondecreasing in $\theta$ for each $x\ge0$.
\end{lemma}

The proof of the lemma is deferred to the Appendix.

The lemma implies that
\begin{equation}\label{lemma-eq}
\sup_{\eta\in[-7/495,0]}\P_\eta(\overline{L}>x)=\P_0(\overline{L}>x).
\end{equation}
for all $x\ge0$.  The probabilities in \eqref{lemma-eq} are parameterized by $\eta$, the expected gain from a one-unit pass-line bet.  In terms of the original $\theta$ parameterization, this would be 
\begin{equation*}
\sup_{\theta\in[0,\theta_0]}\P_\theta(\overline{L}>x)=\P_{\theta_0}(\overline{L}>x)
\end{equation*}
for all $x\ge0$, where $\theta_0$ is the break-even point, the point at which the pass-line bet becomes fair.  In the S--S model, $\theta_0\approx0.0445299$, and in the W--S model, $\theta_0\approx0.0313088$.

Let $\alpha$ be the desired significance level, and let
\begin{equation*}
\overline{L}>\E_0[L]+\Phi^{-1}(1-\alpha)\frac{\text{SD}_0(L)}{\sqrt{n}}.
\end{equation*}
be the critical region.  If $\eta\in[-7/495,0]$, then
\begin{align*}
&\P_\eta(\overline{L}>\E_0[L]+\Phi^{-1}(1-\alpha)\SD_0(L)/\sqrt{n}) \\
&\qquad\le\P_0(\overline{L}>\E_0[L]+\Phi^{-1}(1-\alpha)\SD_0(L)/\sqrt{n}) \\
&\qquad=\P_0\bigg(\frac{\overline{L}-\E_0[L]}{\SD_0(L)/\sqrt{n}}>\Phi^{-1}(1-\alpha)\bigg) \\
&\qquad\approx1-\Phi(\Phi^{-1}(1-\alpha))=\alpha.
\end{align*}
The power of the test under the alternative $\eta$ is then
\begin{align*}
\beta(\eta)&=\P_\eta(\overline{L}>\E_0[L]+\Phi^{-1}(1-\alpha)\SD_0(L)/\sqrt{n}) \\
&=\P_\eta\bigg(\frac{\overline{L}-\E_\eta[L]}{\SD_\eta(L)/\sqrt{n}}>\frac{\E_0[L]-\E_\eta[L]+\Phi^{-1}(1-\alpha)\text{SD}_0(L)/\sqrt{n}}{\text{SD}_\eta(L)/\sqrt{n}}\bigg) \\
&\approx1-\Phi\bigg(\frac{\sqrt{n}(\E_0[L]-\E_\eta[L])+\Phi^{-1}(1-\alpha)\text{SD}_0(L)}{\text{SD}_\eta(L)}\bigg).
\end{align*}

An advantage of basing a test of dice control on observations of the length of the shooter's hand is that data can be collected during the course of a game of craps.  The same is true of our other tests, based on the sample proportion of 7s or the sample proportion of pass-line wins, but we must realize that data for these tests will arrive at different rates.  To see this, the distribution of dice totals, under the null hypothesis of no control, is
\begin{equation*}
p(x)=\frac{6-|x-7|}{36},\quad x=2,3,4,\ldots,12,
\end{equation*}
and the probability that a pass-line decision ends with a seven-out is
\begin{equation*}
p_\text{7o}:=\sum_{x=4,5,6,8,9,10}p(x)\frac{p(7)}{p(x)+p(7)}=\frac{196}{495}.
\end{equation*}
The number of pass-line decisions (hence the number of come-out rolls) in the shooter's hand is geometrically distributed with parameter $p_\text{7o}$, hence its mean is
\begin{equation*}
\gamma:=\frac{1}{p_\text{7o}}=\frac{495}{196}\approx2.52551.
\end{equation*}
Consequently, the the mean number of point rolls in the shooter's hand is
\begin{equation*}
\delta:=\E[L]-\gamma=\frac{1671}{196}-\frac{495}{196}=\frac{1176}{196}=6.
\end{equation*}

Therefore, for each observation of the length of the shooter's hand, we get an average of six observations for the sample proportion of 7s statistic and an average of 2.52551 observations for the sample proportion of pass-line wins statistic.  To compare the power of the $\overline{L}$ test with the power of the test based on the sample proportion of 7s, the sample size of the latter test should be six times as large.  And to compare the power of the $\overline{L}$ test with the power of the test based on the sample proportion of pass-line wins, the sample size of the latter test should be about 2.52551 times as large.  

But there is one flaw in this argument.  Our interest is in comparing power functions, so the null hypothesis of no control is not applicable.  We must recalculate these numbers with $\theta$ dependence.  The probability $p_\text{7o}$ becomes
\begin{align*}
p_\text{7o}(\theta)&:=\sum_{x=4,5,9,10}p_{AA}(x)\frac{p_{AC}(7)}{p_{AC}(x)+p_{AC}(7)}+\sum_{x=6,8}p_{AA}(x)\frac{p_{AB}(7)}{p_{AB}(x)+p_{AB}(7)}\\
&\;=\frac{(4 - \theta) (8 + \theta)(28 - \theta)^2}{144 (10 - \theta) (44 + \theta)} \quad\text{(S--S model)},\\
p_\text{7o}(\theta)&:=\sum_{x=4,5,6,8,9,10}p_{1562}(x)\frac{p_{2424}(7)}{p_{2424}(x)+p_{2424}(7)}\\
&\;=\frac{(3 - 2\,\theta) (196 - 85\,\theta + 6\,\theta^2)}{27 (5 - 2\,\theta) (11 - 3\,\theta)}\quad\text{(W--S model)},
\end{align*}
hence the mean number of pass-line decisions (or the mean number of come-out rolls) in the shooter's hand is
\begin{equation*}
\gamma(\theta):=\frac{1}{p_\text{7o}(\theta)}=\begin{cases}\dfrac{144 (10 - \theta) (44 + \theta)}{(4 - \theta) (8 + \theta)(28 - \theta)^2}&\text{(S--S model)},\\
\noalign{\medskip}
\dfrac{27 (5 - 2\,\theta) (11 - 3\,\theta)}{(3 - 2\,\theta) (196 - 85\,\theta + 6\,\theta^2)}&\text{(W--S model)},\end{cases}
\end{equation*}
and the mean number of point rolls in the shooter's hand is
\begin{equation*}
\delta(\theta):=\E_\theta[L]-\gamma(\theta)=\begin{cases}\dfrac{24}{4-\theta}&\text{(S--S model)},\\
\noalign{\medskip}
\dfrac{18}{3-2\,\theta}&\text{(W--S model)},\end{cases}
\end{equation*}
which is $\rho(\theta)$ in both models, with $\rho$ as in the first reparameterization, hence $\delta(\rho^{-1}(\eta))=\eta$ in both models.

Let us first consider tests using the first reparameterization, in which $\eta$ is the reciprocal of the probability of rolling a 7 on a point roll.  In Table~\ref{tab:powerdata-reparam1} we compare the power of the $\overline{L}$ test under the S--S model with the power of the $\overline{L}$ test under the W--S model and the power of the sample proportion of 7s test under either model.  We assume the simple null hypothesis and a significance level of $\alpha=0.05$.  We consider several values of the alternative $\eta$ and several values of the sample size $n$, keeping in mind that the sample size for the sample proportion of 7s test should be $\eta$ times the sample size of the $\overline{L}$ tests.  (In our approximate power formulas, the sample size does not need to be an integer.)

We find that the $\overline{L}$ test under the W--S model is slightly more powerful than the $\overline{L}$ test under the S--S model, which is slightly more powerful than the sample proportion of 7s test under either model.  For example, for $\eta=6.25$, the power of the $\overline{L}$ test with $n=500$ is 0.2833 (W--S model) and 0.2666 (S--S model), and the power of the sample proportion of 7s test (with $n=500\eta=3125$) is 0.2560 (either model).  The differences are small.

\begin{table}[htb]
\caption{\label{tab:powerdata-reparam1}Approximate power of the $\overline{L}$ test of $H_0:\eta=6$ vs.\ $H_1: \eta>6$ ($\eta$ is the reciprocal of the probability of rolling a 7 on a point roll) for significance level $\alpha=0.05$, various alternatives $\eta$, and various sample sizes $n$, for both models.  Also, approximate power of the sample proportion of 7s test, same scenario.}

\catcode`@=\active\def@{\phantom{0}}
\begin{center}
\begin{tabular}{rrrrrr}
\hline\noalign{\smallskip}
\multicolumn{6}{c}{$\overline{L}$ test, S--S model} \\
\noalign{\smallskip}\hline\noalign{\smallskip}
$n$@ & $\eta=6.125$ & @$\eta=6.25$ & @@$\eta=6.5$ & @@$\eta=7.0$ & @@$\eta=8.0$ \\
\noalign{\smallskip}\hline\noalign{\smallskip}
@100  & 0.0823 & 0.1258 & 0.2452 & 0.5470 & 0.9209 \\
@200  & 0.0967 & 0.1658 & 0.3618 & 0.7701 & 0.9936 \\
@500  & 0.1308 & 0.2666 & 0.6230 & 0.9757 & 1.0000 \\          
1000  & 0.1782 & 0.4086 & 0.8565 & 0.9996 & 1.0000 \\                 
\noalign{\smallskip}\hline\noalign{\smallskip}
\multicolumn{6}{c}{$\overline{L}$ test, W--S model} \\
\noalign{\smallskip}\hline\noalign{\smallskip}
$n$@ & $\eta=6.125$ & @$\eta=6.25$ & @$\eta=6.5$ & @@$\eta=7.0$ & @@$\eta=8.0$ \\
\noalign{\smallskip}\hline\noalign{\smallskip}
@100  & 0.0839 & 0.1302 & 0.2576 & 0.5742 & 0.9358 \\               
@200  & 0.0995 & 0.1735 & 0.3835 & 0.7995 & 0.9960 \\                
@500  & 0.1364 & 0.2833 & 0.6580 & 0.9836 & 1.0000 \\             
1000  & 0.1881 & 0.4368 & 0.8848 & 0.9998 & 1.0000 \\             
\noalign{\smallskip}\hline\noalign{\smallskip}
\multicolumn{6}{c}{Sample proportion of 7s test, either model} \\
\noalign{\smallskip}\hline\noalign{\smallskip}
$n$@ & $\eta=6.125$ & @$\eta=6.25$ & @@$\eta=6.5$ & @@$\eta=7.0$ & @@$\eta=8.0$ \\
\noalign{\smallskip}\hline\noalign{\smallskip}
@100$\eta$  & 0.0762 & 0.1117 & 0.2138 & 0.5192 & 0.9563 \\               
@200$\eta$  & 0.0907 & 0.1517 & 0.3380 & 0.7864 & 0.9993 \\                
@500$\eta$  & 0.1252 & 0.2560 & 0.6280 & 0.9885 & 1.0000 \\             
1000$\eta$  & 0.1740 & 0.4073 & 0.8781 & 1.0000 & 1.0000 \\                    
\noalign{\smallskip}\hline
\end{tabular}
\end{center}
\end{table}

Let us next consider tests using the second reparameterization, in which $\eta$ is the expected gain from a one-unit pass-line bet. We first test the simple null hypothesis $H_0:\eta=-7/495$ versus the alternative hypothesis $H_1:\eta>-7/495$.  In Table~\ref{tab:powerdata-reparam2} we compare the power of the $\overline{L}$ test under the S--S model, the power of the $\overline{L}$ test under the W--S model, the power of the sample proportion of pass-line wins test under the S--S model, and the power of the sample proportion of pass-line wins test under the W--S model.  The significance level is $\alpha=0.05$.  We consider several values of the alternative $\eta$ and several values of the sample size $n$, keeping in mind that the sample size for the sample proportion of pass-line wins test should be $\gamma(\rho^{-1}(\eta))$ times the sample size of the $\overline{L}$ tests.  

We find that, under the S--S model, the $\overline{L}$ test is less powerful than the sample proportion of pass-line wins test.  But under the W--S model, the opposite is true, though to a lesser extent:  The $\overline{L}$ test is more powerful than the sample proportion of pass-line wins test.  For example, under the S--S model, for $\eta=0.025$, the power of the $\overline{L}$ test with $n=500$ is 0.1945, while the power of the sample proportion of pass-line wins test (with $n=\gamma(\rho^{-1}(\eta))\approx1282$) is 0.4038.  Under the W--S model, for $\eta=0.025$, the power of the $\overline{L}$ test with $n=500$ is 0.4554, while the power of the sample proportion of pass-line wins test (with $n=\gamma(\rho^{-1}(\eta))\approx1312$) is 0.4101.  

Next, we test the composite null hypothesis $H_0:\eta\in[-7/495,0]$ versus the alternative hypothesis $H_1:\eta\in(0,13/40)$.  In Table~\ref{tab:powerdata-reparam2-composite} we make the same comparisons as in Table~\ref{tab:powerdata-reparam2}.  The significance level is $\alpha=0.05$.  We consider several values of the alternative $\eta$ and several values of the sample size $n$, keeping in mind that the sample size for the sample proportion of pass-line wins test should be $\gamma(\rho^{-1}(\eta))$ times the sample size of the $\overline{L}$ tests.

Here we find the same results qualitatively as for the simple null hypothesis, although power is reduced, as might be expected.

\begin{table}[H]
\caption{\label{tab:powerdata-reparam2}Approximate power of the $\overline{L}$ test of $H_0:\eta=-7/495$ vs.\ $H_1: \eta>-7/495$ ($\eta$ is the expected gain from a one-unit pass-line bet) for significance level $\alpha=0.05$, various alternatives $\eta$, and various sample sizes $n$, for both models.  Also, approximate power of the sample proportion of pass-line wins test, same scenario.}
\catcode`@=\active\def@{\phantom{0}}
\begin{center}
\begin{tabular}{rrrrrr}
\hline\noalign{\smallskip}
\multicolumn{6}{c}{$\overline{L}$ test, S--S model} \\
\noalign{\smallskip}\hline\noalign{\smallskip}
$n$@ & @@$\eta=0.0$ & $\eta=0.025$ & @$\eta=0.05$ & @@$\eta=0.1$ & @@$\eta=0.2$ \\
\noalign{\smallskip}\hline\noalign{\smallskip}
@100  & 0.0661  &  0.1034 & 0.1532 & 0.2899 & 0.6358 \\
@200  & 0.0726  &  0.1296 & 0.2106 & 0.4326 & 0.8530 \\
@500  & 0.0869  &  0.1945 & 0.3556 & 0.7231 & 0.9927 \\          
1000  & 0.1056  &  0.2874 & 0.5473 & 0.9266 & 1.0000 \\                 
\noalign{\smallskip}\hline\noalign{\smallskip}
\multicolumn{6}{c}{$\overline{L}$ test, W--S model} \\
\noalign{\smallskip}\hline\noalign{\smallskip}
$n$@ & $\eta=0.0$ & $\eta=0.025$ & $\eta=0.05$ & $\eta=0.1$ & $\eta=0.2$ \\
\noalign{\smallskip}\hline\noalign{\smallskip}
@100  & 0.0848 & 0.1835 & 0.3291 & 0.6715 & 0.9816 \\               
@200  & 0.1009 & 0.2620 & 0.4946 & 0.8827 & 0.9997 \\                
@500  & 0.1391 & 0.4554 & 0.7987 & 0.9962 & 1.0000 \\             
1000  & 0.1927 & 0.6844 & 0.9637 & 1.0000 & 1.0000 \\             
\noalign{\smallskip}\hline\noalign{\smallskip}
\multicolumn{6}{c}{sample proportion of pass-line wins test, S--S model} \\
\noalign{\smallskip}\hline\noalign{\smallskip}
$n$@ & @@$\eta=0.0$ & $\eta=0.025$ & @$\eta=0.05$ & @@$\eta=0.1$ & @@$\eta=0.2$ \\
\noalign{\smallskip}\hline\noalign{\smallskip}
@100$\gamma(\rho^{-1}(\eta))$  & 0.0779 & 0.1542 & 0.2698 & 0.5843 & 0.9750 \\
@200$\gamma(\rho^{-1}(\eta))$  & 0.0924 & 0.2240 & 0.4265 & 0.8379 & 0.9997 \\
@500$\gamma(\rho^{-1}(\eta))$  & 0.1269 & 0.4038 & 0.7467 & 0.9941 & 1.0000 \\          
1000$\gamma(\rho^{-1}(\eta))$  & 0.1756 & 0.6320 & 0.9475 & 1.0000 & 1.0000 \\                 
\noalign{\smallskip}\hline\noalign{\smallskip}
\multicolumn{6}{c}{sample proportion of pass-line wins test, W--S model} \\
\noalign{\smallskip}\hline\noalign{\smallskip}
$n$@ & @@$\eta=0.0$ & $\eta=0.025$ & @$\eta=0.05$ & @@$\eta=0.1$ & @@$\eta=0.2$ \\
\noalign{\smallskip}\hline\noalign{\smallskip}
@100$\gamma(\rho^{-1}(\eta))$  & 0.0780 & 0.1560 & 0.2764 & 0.6099 & 0.9869 \\
@200$\gamma(\rho^{-1}(\eta))$  & 0.0926 & 0.2271 & 0.4376 & 0.8598 & 0.9999 \\
@500$\gamma(\rho^{-1}(\eta))$  & 0.1274 & 0.4101 & 0.7607 & 0.9962 & 1.0000 \\          
1000$\gamma(\rho^{-1}(\eta))$  & 0.1764 & 0.6406 & 0.9539 & 1.0000 & 1.0000 \\                 
\noalign{\smallskip}\hline
\end{tabular}
\end{center}
\end{table}

\begin{table}[H]
\caption{\label{tab:powerdata-reparam2-composite}Approximate power of the $\overline{L}$ test of $H_0:\eta\le0$ vs.\ $H_1: \eta>0$ ($\eta$ is the expected gain from a one-unit pass-line bet) for significance level $\alpha=0.05$, various alternatives $\eta$, and various sample sizes $n$, for both models.  Also, approximate power of the sample proportion of pass-line wins test, same scenario.}
\catcode`@=\active\def@{\phantom{0}}
\begin{center}
\begin{tabular}{rrrrrr}
\hline\noalign{\smallskip}
\multicolumn{6}{c}{$\overline{L}$ test, S--S model} \\
\noalign{\smallskip}\hline\noalign{\smallskip}
$n$@ & @@$\eta=0.0$ & $\eta=0.025$ & @$\eta=0.05$ & @@$\eta=0.1$ & @@$\eta=0.2$ \\
\noalign{\smallskip}\hline\noalign{\smallskip}
@100  & 0.0500 & 0.0810 & 0.1239 & 0.2480 & 0.5916 \\
@200  & 0.0500 & 0.0948 & 0.1626 & 0.3664 & 0.8141 \\
@500  & 0.0500 & 0.1271 & 0.2603 & 0.6301 & 0.9863  \\          
1000  & 0.0500 & 0.1719 & 0.3982 & 0.8624 & 0.9999  \\                 
\noalign{\smallskip}\hline\noalign{\smallskip}
\multicolumn{6}{c}{$\overline{L}$ test, W--S model} \\
\noalign{\smallskip}\hline\noalign{\smallskip}
$n$@ & $\eta=0.0$ & $\eta=0.025$ & $\eta=0.05$ & $\eta=0.1$ & $\eta=0.2$ \\
\noalign{\smallskip}\hline\noalign{\smallskip}
@100  & 0.0500 & 0.1225 & 0.2445 & 0.5852 & 0.9712 \\               
@200  & 0.0500 & 0.1609 & 0.3623 & 0.8101 & 0.9992 \\                
@500  & 0.0500 & 0.2580 & 0.6261 & 0.9859 & 1.0000 \\             
1000  & 0.0500 & 0.3954 & 0.8602 & 0.9999 & 1.0000 \\             
\noalign{\smallskip}\hline\noalign{\smallskip}
\multicolumn{6}{c}{sample proportion of pass-line wins test, S--S model} \\
\noalign{\smallskip}\hline\noalign{\smallskip}
$n$@ & @@$\eta=0.0$ & $\eta=0.025$ & @$\eta=0.05$ & @@$\eta=0.1$ & @@$\eta=0.2$ \\
\noalign{\smallskip}\hline\noalign{\smallskip}
@100$\gamma(\rho^{-1}(\eta))$  & 0.0500 & 0.1066 & 0.2001 & 0.4926 & 0.9572 \\
@200$\gamma(\rho^{-1}(\eta))$  & 0.0500 & 0.1402 & 0.3059 & 0.7449 & 0.9991 \\
@500$\gamma(\rho^{-1}(\eta))$  & 0.0500 & 0.2266 & 0.5614 & 0.9774 & 1.0000 \\          
1000$\gamma(\rho^{-1}(\eta))$  & 0.0500 & 0.3523 & 0.8161 & 0.9998 & 1.0000 \\                 
\noalign{\smallskip}\hline\noalign{\smallskip}
\multicolumn{6}{c}{sample proportion of pass-line wins test, W--S model} \\
\noalign{\smallskip}\hline\noalign{\smallskip}
$n$@ & @@$\eta=0.0$ & $\eta=0.025$ & @$\eta=0.05$ & @@$\eta=0.1$ & @@$\eta=0.2$ \\
\noalign{\smallskip}\hline\noalign{\smallskip}
@100$\gamma(\rho^{-1}(\eta))$  & 0.0500 & 0.1074 & 0.2044 & 0.5158 & 0.9754 \\
@200$\gamma(\rho^{-1}(\eta))$  & 0.0500 & 0.1417 & 0.3136 & 0.7706 & 0.9997 \\
@500$\gamma(\rho^{-1}(\eta))$  & 0.0500 & 0.2297 & 0.5750 & 0.9835 & 1.0000 \\          
1000$\gamma(\rho^{-1}(\eta))$  & 0.0500 & 0.3577 & 0.8289 & 0.9999 & 1.0000 \\                 
\noalign{\smallskip}\hline
\end{tabular}
\end{center}
\end{table}

\section{The distribution of $L$}\label{sec:distrib-L}

Assuming fair dice and no control, Ethier and Hoppe (2010) found an explicit formula for the distribution of $L$, the length of the shooter's hand, as a linear combination of four geometric distributions.  More specifically,
\begin{align*}
f(x):&=\P_0(L=x)\\
&\;=\sum_{i=1}^4 c_i\, e_i^{x-1}(1-e_i)=\sum_{i=1}^4 c_i\,\text{GEOM}[1-e_i](x),\qquad x\in\{2,3,4,\ldots\},
\end{align*}
where
\begin{align*}
e_1:=e(1,1)&\approx0.862473751659322030,\\
e_2:=e(1,-1)&\approx0.741708271459795977,\\
e_3:=e(-1,1)&\approx0.709206775794379015,\\
e_4:=e(-1,-1)&\approx0.186611201086502979,
\end{align*}
are the non-unit eigenvalues of the stochastic matrix 
\begin{equation}\label{P}
\bm P_0:=\frac{1}{36}\begin{pmatrix}12&6&8&10&0\\3&27&0&0&6\\4&0&26&0&6\\5&0&0&25&6\\0&0&0&0&36\end{pmatrix},
\end{equation}
and
\begin{align*}
e(u,v):=\frac58+\frac{u}{72}\sqrt{\frac{349+\alpha}{3}}+\frac{v}{72}\sqrt{\frac{698-\alpha}{3}-2136\,u\sqrt{\frac{3}{349+\alpha}}}
\end{align*}
with
\begin{equation*}
\alpha:=2\sqrt{9829}\cos\bigg[\frac13\cos^{-1}\bigg(\!\!-\frac{710369}{9829\sqrt{9829}}\bigg)\bigg].
\end{equation*}
Furthermore,
\begin{align*}
c_1&:=c(e_1,e_2,e_3,e_4)\approx\phantom{-}1.211844812464518572,\\
c_2&:=c(e_2,e_3,e_4,e_1)\approx-0.006375542263784777,\\
c_3&:=c(e_3,e_4,e_1,e_2)\approx-0.004042671248651503,\\
c_4&:=c(e_4,e_1,e_2,e_3)\approx-0.201426598952082292,
\end{align*}
are the coefficients of the linear combination, where
\begin{align*}
c(x_1,x_2,x_3,x_4)&:=(-25+36\,x_1)[4835-5580(x_2+x_3+x_4)\\
&\qquad\quad{}+6480(x_2 x_3+x_2 x_4+x_3 x_4)-7776\,x_2 x_3 x_4]\\
&\qquad{}/[38880(x_1-x_2)(x_1-x_3)(x_1-x_4)].
\end{align*}
Incidentally, a linear combination of probability distributions is not necessarily a probability distribution, but in this case it is, because
\begin{equation*}
c_1+c_2+c_3+c_4=1\quad\text{and}\quad c_1\,e_1+c_2\,e_2+c_3\,e_3+c_4\,e_4=1.
\end{equation*}

In this section we generalize that result from the fair with no control model to the S--S and W--S dice-control models.

The first step is to generalize the five-state Markov chain with transition matrix $\bm P_0$ as in \eqref{P} used by Ethier and Hoppe (2010).    This has already been done in \eqref{SS-P_theta} (S--S model) and \eqref{WS-P_theta} (W--S model).   

\begin{theorem}
Under either the S--S model or the W--S model, the distribution of $L$, the length of the shooter's hand, has the form
\begin{align}\label{f(x,theta)}
f(x,\theta)&:=\P_\theta(L=x)\nonumber\\
&\;=\sum_{i=1}^4 c_i(\theta)\,e_i(\theta)^{x-1}(1-e_i(\theta))\nonumber\\
&\;=\sum_{i=1}^4 c_i(\theta)\,{\rm GEOM}[1-e_i(\theta)](x),\qquad x\in\{2,3,4,\ldots\},
\end{align}
where $1>e_1(\theta)>e_2(\theta)>e_3(\theta)>e_4(\theta)>0$ are the non-unit eigenvalues of $\bm P_\theta$, and $c_1(\theta)>0$, $c_2(\theta)<0$, $c_3(\theta)<0$, and $c_4(\theta)<0$ are suitably defined functions satisfying
\begin{equation*}
c_1(\theta)+c_2(\theta)+c_3(\theta)+c_4(\theta)=1
\end{equation*}
and
\begin{equation*}
c_1(\theta)\,e_1(\theta)+c_2(\theta)\,e_2(\theta)+c_3(\theta)\,e_3(\theta)+c_4(\theta)\,e_4(\theta)=1.
\end{equation*}

More precisely, under the S--S model, let
\begin{align*}
q(\theta)&:= 727417856 + 1090622976\,\theta + 592227264\,\theta^2 + 146776064\,\theta^3 \\
         &\qquad{}+ 18260400\,\theta^4 + 1567554\,\theta^5 + 85295\,\theta^6, \\
r(\theta)&:=2(314528 + 263680\,\theta + 74334\,\theta^2 + 9592\,\theta^3 + 527\,\theta^4), \\
s(\theta)&:= (8 + \theta) (22784 + 13520\,\theta + 2171\,\theta^2), \\
t(\theta)&:= 22336 + 10480\,\theta + 1123\,\theta^2.
\end{align*}
Then
\begin{align}
\begin{split}\label{e_i(theta)}
e_1(\theta)&:=e(1,1,\theta),\\  
e_2(\theta)&:=e(1,-1,\theta),\\ 
e_3(\theta)&:=e(-1,1,\theta),\\  
e_4(\theta)&:=e(-1,-1,\theta),
\end{split}
\end{align}
where
\begin{align}\label{e(u,v,theta)}
e(u,v,\theta)&:=\frac58-\frac{\theta}{64}+\frac{u}{576}\sqrt{\frac{t(\theta)+8\,\alpha(\theta)}{3}}\nonumber\\
\noalign{\smallskip}
&\qquad{}+\frac{v}{576}\sqrt{\frac{2\,t(\theta)-8\,\alpha(\theta)}{3}-6\,u\,s(\theta)\sqrt{\frac{3}{t(\theta)+8\,\alpha(\theta)}}}\quad
\end{align}
with 
\begin{equation*}
\alpha(\theta):=2\sqrt{r(\theta)}\cos\bigg[\frac13\cos^{-1}\bigg(\!\!-\frac{q(\theta)}{2\,r(\theta)^{3/2}}\bigg)\bigg].
\end{equation*}
Finally, with
\begin{align*}
&c(e_1,e_2,e_3,e_4,\theta)\\
&\qquad{}:=\frac{(3-4\,e_1)(26+\theta-36\,e_1)(100-\theta-144\,e_1)\prod_{i=2}^4 (20+\theta-24\,e_i)}{1152(2+\theta)(8+\theta)(20+7\,\theta)(e_2-e_1)(e_3-e_1)(e_4-e_1)},
\end{align*}
we have
\begin{align*}
c_1(\theta)&:=c(e_1(\theta),e_2(\theta),e_3(\theta),e_4(\theta),\theta),\\
c_2(\theta)&:=c(e_2(\theta),e_3(\theta),e_4(\theta),e_1(\theta),\theta),\\
c_3(\theta)&:=c(e_3(\theta),e_4(\theta),e_1(\theta),e_2(\theta),\theta),\\
c_4(\theta)&:=c(e_4(\theta),e_1(\theta),e_2(\theta),e_3(\theta),\theta).
\end{align*}

And under the W--S model, let
\begin{align*}
q(\theta)&:= 710369 + 258462\,\theta - 63957\,\theta^2 - 5832\,\theta^3 + 6027\,\theta^4 + 522\,\theta^5 - 55\,\theta^6, \\
r(\theta)&:= 9829 + 1356\,\theta - 238\,\theta^2 + 12\,\theta^3 + \theta^4, \\
s(\theta)&:= 1068 + 246\,\theta + 5\,\theta^2 - 2\,\theta^3, \\
t(\theta)&:= 349 + 48\,\theta - 2\,\theta^2.
\end{align*}
Then \eqref{e_i(theta)} holds, where
\begin{align*}
e(u,v,\theta)&:=\frac58+\frac{\theta}{12}+\frac{u}{72}\sqrt{\frac{t(\theta)+\alpha(\theta)}{3}}\nonumber\\
\noalign{\smallskip}
&\qquad{}+\frac{v}{72}\sqrt{\frac{2\,t(\theta)-\alpha(\theta)}{3}-2\,u\,s(\theta)\sqrt{\frac{3}{t(\theta)+\alpha(\theta)}}}\quad
\end{align*}
with
\begin{equation*}
\alpha(\theta):=2\sqrt{r(\theta)}\cos\bigg[\frac13\cos^{-1}\bigg(\!\!-\frac{q(\theta)}{r(\theta)^{3/2}}\bigg)\bigg].
\end{equation*}
Finally, with
\begin{align*}
&c(e_1,e_2,e_3,e_4,\theta)\\
&\qquad{}:=\Big[69984(9+\theta-12\,e_1)(13+2\,\theta-18\,e_1)(25+3\,\theta-36\,e_1)\\
&\qquad\qquad{}\cdot(e_2-e_3)(e_2-e_4)(e_3-e_4)\prod_{i=2}^4 (15+2\,\theta-18\,e_i)\Big]/d(\theta)
\end{align*}
and
\begin{align*}
d(\theta)
&:=(3 + \theta) (5 + \theta) [544195584 (e_1(\theta)^2 e_2(\theta)^3 e_3(\theta) + e_1(\theta)^3 e_2(\theta) e_3(\theta)^2 \\
&\qquad{}+ e_1(\theta) e_2(\theta)^2 e_3(\theta)^3 + e_1(\theta)^3 e_2(\theta)^2 e_4(\theta) + e_1(\theta)^2 e_2(\theta) e_4(\theta)^3 \\
&\qquad{}+ e_1(\theta) e_2(\theta)^3 e_4(\theta)^2 + e_1(\theta)^2 e_3(\theta)^3 e_4(\theta) + e_1(\theta)^3 e_3(\theta) e_4(\theta)^2 \\
&\qquad{}+ e_1(\theta) e_3(\theta)^2 e_4(\theta)^3 + e_2(\theta)^3 e_3(\theta)^2 e_4(\theta) + e_2(\theta)^2 e_3(\theta) e_4(\theta)^3 \\
&\qquad{}+ e_2(\theta) e_3(\theta)^3 e_4(\theta)^2) - (448331256 + 349805385\,\theta + 113332133\,\theta^2 \\
&\qquad{}+ 19480471\,\theta^3 + 1870151\,\theta^4 + 94864\,\theta^5 + 1980\,\theta^6)],
\end{align*}
we have 
\begin{align*}
c_1(\theta)&:=c(e_1(\theta),e_2(\theta),e_3(\theta),e_4(\theta),\theta),\\
c_2(\theta)&:=-c(e_2(\theta),e_3(\theta),e_4(\theta),e_1(\theta),\theta),\\
c_3(\theta)&:=c(e_3(\theta),e_4(\theta),e_1(\theta),e_2(\theta),\theta),\\
c_4(\theta)&:=-c(e_4(\theta),e_1(\theta),e_2(\theta),e_3(\theta),\theta).
\end{align*}
\end{theorem}

The proof is deferred to the Appendix.  Notice that the coefficients $c_1(\theta)$, $c_2(\theta)$, $c_3(\theta)$, and $c_4(\theta)$ are a little more complicated in the W--S model than in the S--S model.

It is worth mentioning that the tail probabilities have the form
\begin{equation}\label{tail-eigen}
t(x,\theta):=\P_\theta(L\ge x)=\sum_{i=1}^4 c_i(\theta)\,e_i(\theta)^{x-1},\qquad x\in\{2,3,4,\ldots\}.
\end{equation}

To illustrate the theorem, we evaluate $t(154,\theta)$ in \eqref{tail-eigen} for three values of $\theta$ corresponding to probabilities 1/6, 1/7, and 1/8 of rolling a 7 on a point roll.  Under the S--S model,
\begin{equation*}
\frac{1}{t(154,0)}\approx5.590\,\text{B},\quad \frac{1}{t(154,4/7)}\approx148.8\,\text{M},\quad \frac{1}{t(154,1)}\approx11.31\,\text{M},
\end{equation*}
and under the W--S model.
\begin{equation*}
\frac{1}{t(154,0)}\approx5.590\,\text{B},\quad \frac{1}{t(154,3/14)}\approx167.9\,\text{M},\quad \frac{1}{t(154,3/8)}\approx13.07\,\text{M}.
\end{equation*}
In words, there is one chance in 5.59 billion that a random shooter will achieve a hand of 154 or more rolls, as did Patricia DeMauro on May 23, 2009, in Atlantic City (Ethier and Hoppe, 2010).  A skilled shooter who can reduce the probability of rolling a 7 on a point roll to 1/8 increases his or her probability of a hand of 154 or more rolls to one chance in 11.31 million under the S--S model and to one chance in 13.07 million under the W--S model.

\section{The likelihood ratio test}\label{sec:LRT}

For a random sample of size $n$ from the distribution \eqref{f(x,theta)}, denoted by $\bm x=(x_1,x_2,\ldots,x_n)$, let 
\begin{equation}\label{likelihood-function}
\mathscr{L}(\theta,\bm x):=\prod_{i=1}^n f(x_i,\theta)
\end{equation}
be the likelihood function of the sample.  Define the \textit{likelihood ratio statistic} for testing $H_0:\theta=0$ vs.\ $H_1:\theta>0$ by
\begin{equation*}
\Lambda_n(\bm x)=\frac{\mathscr{L}(0,\bm x)}{\max_{\theta\in[0,1]}\mathscr{L}(\theta,\bm x)}=\frac{\mathscr{L}(0,\bm x)}{\mathscr{L}(\hat\theta(\bm x),\bm x)},
\end{equation*}
where $\hat\theta(\bm x):=\text{arg\,max}_{\theta\in[0,1]}\mathscr{L}(\theta,\bm x)$ is the \textit{maximum likelihood estimator} of $\theta$.
A well-known theorem (e.g., Theorem 21.1 of DasGupta, 2008) tells us that, under $H_0$,
\begin{equation*}
-2\log \Lambda_n(\bm x)\to_d\chi^2(1) \text{ as } n\to\infty.
\end{equation*}
Thus, we reject $H_0$ at the approximate significance level $\alpha$ if 
\begin{equation*}
-2\log\Lambda_n(\bm x)>[\Phi^{-1}(1-\alpha/2)]^2.
\end{equation*}

There appears to be no analytical formula for the maximum likelihood estimator in this problem (for either model).  Therefore we use simulation to estimate the power at various alternatives.  We restrict attention to samples of size $n=500$ and significance level $\alpha=0.05$.

First, to simulate a random sample from $f(\,\cdot\,,\theta)$, we run the Markov chain with initial state co and transition matrix $\bm P_\theta$ as in \eqref{SS-P_theta} (S--S model) or \eqref{WS-P_theta} (W--S model) until absorption, $n$ times.  Under the S--S model, dice totals are simulated from $p_{AA}(\,\cdot\,,\theta)$ using
\begin{equation*}
\text{total}=\begin{cases}(\lfloor{6U_1}\rfloor+1)+(\lfloor{6U_2}\rfloor+1)&\text{if $U_0<1-\theta$},\\
(\lfloor{4U_1}\rfloor+2)+(\lfloor{4U_2}\rfloor+2)&\text{otherwise},\end{cases}
\end{equation*} 
from $p_{AB}(\,\cdot\,,\theta)$ using
\begin{equation*}
\text{total}=\begin{cases}(\lfloor{6U_1}\rfloor+1)+(\lfloor{6U_2}\rfloor+1)&\text{if $U_0<1-\theta$},\\
(\lfloor{4U_1}\rfloor+2)+(2\lfloor{4U_2}\rfloor+1)&\text{if $U_2<1/2$ and $U_0\ge1-\theta$},\\
(\lfloor{4U_1}\rfloor+2)+(2\lfloor{4U_2}\rfloor)&\text{if $U_2\ge1/2$ and $U_0\ge1-\theta$},\end{cases}
\end{equation*} 
and from $p_{AC}(\,\cdot\,,\theta)$ using
\begin{equation*}
\text{total}=\begin{cases}(\lfloor{6U_1}\rfloor+1)+(\lfloor{6U_2}\rfloor+1)&\text{if $U_0<1-\theta$},\\
(\lfloor{4U_1}\rfloor+2)+(\lfloor{4U_2}\rfloor+1)&\text{if $U_2<1/2$ and $U_0\ge1-\theta$},\\
(\lfloor{4U_1}\rfloor+2)+(\lfloor{4U_2}\rfloor+3)&\text{if $U_2\ge1/2$ and $U_0\ge1-\theta$}.\end{cases}
\end{equation*} 
Under the W--S model, dice totals are simulated from $p_{1562}(\,\cdot\,,\theta)$ using
\begin{align*}
\text{pre-total}&=(\lfloor{6U_1}\rfloor+1)+(\lfloor{6U_2}\rfloor+1); \\
\text{total}&=\begin{cases}7&\text{if $|\text{pre-total}-7|=5$ and $U_0\ge1-\theta$, or}\\
&\text{if $|\text{pre-total}-7|=3$ and $U_3<1/3$ and $U_0\ge1-\theta$},\\
\text{pre-total}&\text{otherwise},\end{cases}
\end{align*}
and from $p_{2424}(\,\cdot\,,\theta)$ using
\begin{align*}
\text{pre-total}&=(\lfloor{6U_1}\rfloor+1)+(\lfloor{6U_2}\rfloor+1); \\
\text{total}&=\begin{cases}2(\lfloor4U_4\rfloor+2)&\text{if $\text{pre-total}=7$ and $U_3<2/3$ and $U_0\ge1-\theta$},\\
\text{pre-total}&\text{otherwise}.\end{cases}
\end{align*}
In the above formulas, $U_0,U_1,U_2,\ldots$ are i.i.d UNIF $[0,1]$ obtained by successive calls to a random number generator.

To evaluate the LR statistic $\Lambda_n(\bm x)$ numerically requires maximizing the very complicated likelihood function \eqref{likelihood-function}.  The formulation
\begin{equation*}
\mathscr{L}(\theta,\bm x):=\prod_{j=2}^{\max x_i} f(j,\theta)^{|\{1\le i\le n: \; x_i=j\}|}
\end{equation*}
is more efficient, reducing the number of calls to $f(\,\cdot\,,\theta)$ substantially.  So we use the \texttt{NMaximize} command in \textit{Mathematica}, which maximizes a function numerically.  (To prevent underflow, we multiply each $f(j,\theta)$ by 20.)  

\begin{figure}[htb]
\begin{center}
\includegraphics[width=3.in]{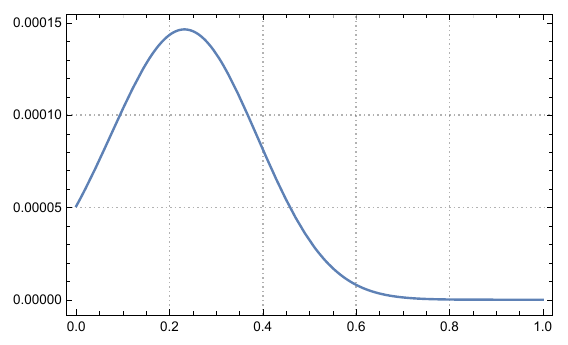}
\end{center}
\caption{\label{fig:likelihood-example}A plot of the likelihood function for the sample \eqref{sample-example}, multiplied by a constant.  Here the data were generated using $\theta = 0.199803$, and the maximum likelihood estimator is $\hat\theta(\bm x)\approx0.231991$.}
\end{figure}

To make this more transparent, we consider an example.  Under the S--S model, suppose we want to estimate the power of the likelihood ratio test of $H_0:\eta=-7/495$ versus $H_1:\eta>-7/495$ at the alternative $\eta=0.05$.  (Assume sample size $n=500$ and significance level $\alpha=0.05$.)  We use $\theta=0.199803\approx\rho^{-1}(0.05)$ and generate the random sample
\begin{align}
\begin{split}\label{sample-example}
&\{45,64,41,43,30,35,38,31,18,22,22,11,11,15,12,10,7,7,7,\\
&\qquad 2,5,3,1,2,2,2,2,1,1,0,2,0,2,0,2,1,1,1,0,0,0,1\},
\end{split}
\end{align}
where 45 is the number of hands of length 2, 64 is the number of hands of length 3, and so on.  The longest hand in the sample was 43 rolls.
We can plot the resulting likelihood function; see Figure~\ref{fig:likelihood-example}.  We find that the maximum likelihood estimator is $\hat\theta(\bm x)\approx0.231991$, $\mathscr{L}(\hat\theta(\bm x),\bm x)\approx0.000146516$, and $\mathscr{L}(0,\bm x)\approx0.0000507253$.  Thus, the LR statistic is 2.12142, which does not exceed $(1.960)^2$, so we fail to reject $H_0$ for this sample. 

As one might imagine, this maximization takes time.  Therefore, we limit our simulation to 10,000 replications of the random sample of size $n=500$, which suffices for two decimal places of accuracy.  

We can carry out the analysis for both models under the original parameterizations ($\theta\in[0,1]$ in the S--S model, $\theta\in[0,0.674351]$ in the W--S model; here 0.674351 is $\rho^{-1}(13/40$)).  Then we reparameterize by $\eta$, the expected gain from a one-unit pass-line bet.

Table~\ref{tab:power-LR-test} shows that the sample-mean test outperforms the likelihood ratio test significantly, in both models, in the sense that it has higher power at various alternatives.  We limited our analysis to random samples of size $n=500$, significance level $\alpha=0.05$, and the  simple null hypothesis, but undoubtedly this conclusion holds more generally. 

\begin{table}[H]
\caption{\label{tab:power-LR-test}Comparing the power of the $\overline{L}$ test of $H_0:\eta=-7/495$ vs.\ $H_1: \eta>-7/495$ ($\eta$ is the expected gain from a one-unit pass-line bet) with that of the likelihood ratio test.  We assume sample size $n=500$ and significance level $\alpha=0.05$.  Simulation results are based on 10,000 replications of the sample of size $n$.}
\catcode`@=\active\def@{\phantom{0}}
\begin{center}
\begin{tabular}{lccc}
\hline
\noalign{\smallskip}
\multicolumn{4}{c}{S--S model} \\
\noalign{\smallskip}
\hline
\noalign{\smallskip}
        &  & approximate &  simulated power  \\
$\eta$ & @$\theta=\rho^{-1}(\eta)$@  & power of $\overline{L}$ test   & of LR test ($\pm\text{s.e.}$) \\
\noalign{\smallskip}        
\hline
\noalign{\smallskip}
0.0   & 0.0445299 & 0.0869 & 0.0441 ($\pm 0.0021$) \\
0.025 & 0.122583@ & 0.1945 & 0.1156 ($\pm 0.0032$) \\
0.05  & 0.199803@ & 0.3556 & 0.2450 ($\pm 0.0043$) \\
0.1   & 0.351832@ & 0.7231 & 0.6118 ($\pm 0.0049$) \\
0.2   & 0.646824@ & 0.9927 & 0.9868 ($\pm 0.0011$) \\
\noalign{\smallskip}        
\hline
\noalign{\smallskip}
\multicolumn{4}{c}{W--S model} \\
\noalign{\smallskip}
\hline
\noalign{\smallskip}
         &  & approximate  & simulated power \\
$\eta$ & $\theta=\rho^{-1}(\eta)$ & power of $\overline{L}$ test &  of LR test $(\pm\text{s.e.})$ \\
\noalign{\smallskip}        
\hline
\noalign{\smallskip}
0.0   & 0.0313088 & 0.1391 & 0.0733 ($\pm 0.0026$) \\
0.025 & 0.0859974 & 0.4554 & 0.3206 ($\pm 0.0047$) \\
0.05  & 0.139835@ & 0.7987 & 0.6855 ($\pm 0.0046$) \\
0.1   & 0.244931@ & 0.9962 & 0.9932 ($\pm 0.0008$) \\
0.2   & 0.444642@ & 1.0000 & 1.0000 ($\pm 0.0000$) \\
\noalign{\smallskip}        
\hline
\end{tabular}
\end{center}
\end{table}

\section{Conclusions}\label{conclusions}

The question of whether dice control is possible at casino craps is controversial and remains unsettled.  If there are people who have the ability to control the dice in a way sufficient to gain an advantage at craps, it should be possible to establish it at an acceptable level of statistical significance.  In this paper we have described two statistical models for dice control, the Smith--Scott model and the Wong--Shackleford model.  We have  proposed two reparameterizations of the models, in terms of the reciprocal of the probability of rolling a 7 on a point roll or, better yet, the expected gain from a one-unit pass-line bet.  There are two null hypotheses of interest, the simple hypothesis of no control and the composite hypothesis of insufficient control to win at craps.  We have proposed four relevant test statistics: (a) the sample proportion of 7s; (b) the sample proportion of pass-line wins; (c) the sample mean of hand-length observations; and (d) the likelihood ratio statistic for a hand-length sample.  We have approximated the power for various tests at various alternatives and various sample sizes.  

Our first conclusion is that tests based on $\overline{L}$, the sample mean of hand-length observations, are significantly more powerful than the corresponding likelihood ratio tests, so the likelihood ratio statistic can be dismissed as a viable test statistic.

Our second conclusion is less definitive.  Using the first reparameterization and the simple null hypothesis, the $\overline{L}$ test under the W--S model is slightly more powerful than the $\overline{L}$ test under the S--S model, which is slightly more powerful than the sample proportion of 7s test under either model.  But the values of the power functions are close enough that none of the three tests should be ruled out or in solely on this basis.

Our third conclusion concerns the second reparameterization and either of the two null hypotheses.  Under the S--S model, the $\overline{L}$ test is less powerful than the sample proportion of pass-line wins test.  But under the W--S model, the opposite is true, though to a lesser extent:  the $\overline{L}$ test is more powerful than the sample proportion of pass-line wins test.  And, although it is not important, the $\overline{L}$ test under the W--S model is more powerful than the $\overline{L}$ test under the S--S model.  The reason this is not important is that model choice should not depend on power functions but rather on which model best fits the data to be collected.  Is the shooter attempting to achieve on-axis control or control by correlation?  A preliminary sample may need to be examined to determine which model is most appropriate, or whether a third model might better describe the data.

Finally, for a test of dice control to have credibility, the data collection must be witnessed by independent observers, as was done in the 2004 Wong experiment mentioned in the Introduction.  Data could come in the form of a hand-length sample, but more useful would be a sample of dice totals, from which could be inferred the sample mean of hand lengths, the sample proportion of pass-line wins, and the sample proportion of 7s on point rolls.  More useful still would be a sample of dice rolls (e.g., instead of recording a 7, one would record $(1,6)$ or $(2,5)$ or $(3,4)$).  This would help to determine how well the data fit the model.  Needless to say, it would also be useful to know which dice set the shooter is using at each roll.

A significant challenge is obtaining a sufficiently large sample size.  Merging samples from multiple experimenters, as was done recently in a large-scale coin-tossing experiment (Bartoš et al., 2025), would be problematic without a consistent dice control technique by the various shooters.

\section*{Appendix}

\subsection*{Proof of Theorem 2 (case of S--S model)}

From \eqref{SS-P_theta},
\begin{equation*}
\bm P_\theta=\frac{1}{144}\begin{pmatrix}12(4-\theta) & 6(4-\theta) & 4(8+\theta) & 2(20+7\,\theta) & 0 \\
6(2+\theta) & 108 & 0 & 0 & 6(4-\theta) \\
2(8+\theta) & 0 & 4(26+\theta) & 0 & 6(4-\theta) \\
20+7\,\theta & 0 & 0 & 100-\theta & 6(4-\theta) \\
0 & 0 & 0 & 0 & 144\end{pmatrix}.
\end{equation*}
As in the $\theta=0$ case, one can show that the non-unit eigenvalues of $P_\theta$ interlace the diagonal entries (except the first one):
\begin{equation*}
1>e_1(\theta)>\frac{108}{144}>e_2(\theta)>\frac{104+4\,\theta}{144}>e_3(\theta)>\frac{100-\theta}{144}>e_4(\theta)>0.
\end{equation*}
This holds for $0\le\theta<1$, and also for $\theta=1$ except for the third and fourth inequalities, which become equalities.
The non-unit eigenvalues are roots of the quartic equation
\begin{equation}\label{quartic}
a(\theta)z^4+b(\theta)z^3+c(\theta)z^2+d(\theta)z+e(\theta)=0,
\end{equation}
where
\begin{align*}
a(\theta)&=2985984, \\
b(\theta)&=-186624 (40 - \theta), \\
c(\theta)&=288 (22904 - 1870\,\theta - 55\,\theta^2), \\
d(\theta)&=-12 (195424 - 40920\,\theta - 2124\,\theta^2 - 19\,\theta^3), \\
e(\theta)&=252800 - 144032\,\theta - 10176\,\theta^2 - 176\,\theta^3 - \theta^4.
\end{align*}
The quartic formula gives complicated expressions for the four roots $e_1(\theta)$, $e_2(\theta)$, $e_3(\theta)$, and $e_4(\theta)$, which can be massaged into the form stated in the theorem. 

Next, \textit{Mathematica} gives five right eigenvectors of $\bm P_\theta$, which can be expressed in terms of $\theta$ and the eigenvalues.  We define the functions
\begin{align*}
r_1(x,\theta)&:=-\frac{100-\theta-144\,x}{20+7\,\theta},\\
r_2(x,\theta)&:=\frac{6(2+\theta)(100-\theta-144\,x)}{(20+7\,\theta)(108-144\,x)},\\
r_3(x,\theta)&:=\frac{2(8+\theta)(100-\theta-144\,x)}{(20+7\,\theta)(104+4\,\theta-144\,x)}, 
\end{align*}
and
\begin{equation*}
\bm r(x,\theta):=\begin{pmatrix}r_1(x,\theta)\\ r_2(x,\theta)\\ r_3(x,\theta)\\ 1\\ 0\end{pmatrix}.
\end{equation*}
Then $(1,1,1,1,1)^\T$, $\bm r(e_1(\theta),\theta)$, $\bm r(e_2(\theta),\theta)$, $\bm r(e_3(\theta),\theta)$, and $\bm r(e_4(\theta),\theta)$ are five right eigenvectors corresponding to the five eigenvalues $1$, $e_1(\theta)$, $e_2(\theta)$, $e_3(\theta)$, and $e_4(\theta)$ of $\bm P_\theta$.  In the special case $\theta=0$, these five right eigenvectors are not the same ones as those used by Ethier and Hoppe (2010).  Define the $5\times5$ matrix $\bm R_\theta$ to be the matrix comprising these five columns.  The rows of the $5\times5$ matrix $\bm L_\theta:=\bm R_\theta^{-1}$ are then left eigenvectors.  We use the spectral representation to derive
\begin{align}\label{tail-formula}
\P_\theta(L\ge x)&=1-[\bm P_\theta^{x-1}]_{\text{co},\text{7o}}\nonumber\\
&=1-[\bm R_\theta\,\text{diag}(1,e_1(\theta)^{x-1},e_2(\theta)^{x-1},e_3(\theta)^{x-1},e_4(\theta)^{x-1})\bm L_\theta]_{\text{co},\text{7o}}\nonumber\\
&=c_1(\theta)\,e_1(\theta)^{x-1}+c_2(\theta)\,e_2(\theta)^{x-1}+c_3(\theta)\,e_3(\theta)^{x-1}+c_4(\theta)\,e_4(\theta)^{x-1}
\end{align}
with help from \textit{Mathematica}, and \eqref{f(x,theta)} follows.

\subsection*{Proof of Theorem 2 (case of W--S model)}

The proof is similar to that in the case of the S--S model, so we highlight only the differences.
From \eqref{WS-P_theta},
\begin{equation*}
\bm P_\theta=\frac{1}{36}\begin{pmatrix}2(6+\theta) & 2(3-\theta) & 8 & 10 & 0 \\
3+\theta & 3(9+\theta) & 0 & 0 & 2(3-2\,\theta) \\
4 & 0 & 2(13+2\,\theta) & 0 & 2(3-2\,\theta) \\
5+\theta & 0 & 0 & 25+3\,\theta & 2(3-2\,\theta) \\
0 & 0 & 0 & 0 & 36\end{pmatrix}.
\end{equation*}
As before, the non-unit eigenvalues satisfy
\begin{equation*}
1>e_1(\theta)>\frac{27+3\,\theta}{36}>e_2(\theta)>\frac{26+4\,\theta}{36}>e_3(\theta)>\frac{25+3\,\theta}{36}>e_4(\theta)>0.
\end{equation*}
This holds for $0\le\theta<1$, and also for $\theta=1$ except for the third and fourth inequalities, which become equalities.  The non-unit eigenvalues are roots of the quartic equation \eqref{quartic}, where
\begin{align*}
a(\theta)&=209952, \\
b(\theta)&=-34992 (15 + 2\,\theta), \\
c(\theta)&=162 (2863 + 786\,\theta + 55\,\theta^2), \\
d(\theta)&=-9 (18321 + 7926\,\theta + 1163\,\theta^2 + 58\,\theta^3), \\
e(\theta)&=17775 + 11457\,\theta + 2768\,\theta^2 + 298\,\theta^3 + 12\,\theta^4.
\end{align*}

Next, for the right eigenvectors of $\bm P_\theta$,  we define 
\begin{align*}
r_1(x,\theta)&:=-\frac{25+3\,\theta-36\,x}{5+\theta},\\
r_2(x,\theta)&:=\frac{(3+\theta)(25+3\,\theta-36\,x)}{(5+\theta)(27+3\,\theta-36\,x)},\\
r_3(x,\theta)&:=\frac{4(25+3\,\theta-36\,x)}{(5+\theta)(26+4\,\theta-36\,x)},
\end{align*}
and
\begin{equation*}
\bm r(x,\theta):=\begin{pmatrix}r_1(x,\theta)\\ r_2(x,\theta)\\ r_3(x,\theta)\\ 1\\ 0\end{pmatrix}.
\end{equation*}
Then $(1,1,1,1,1)^\T$, $\bm r(e_1(\theta),\theta)$, $\bm r(e_2(\theta),\theta)$, $\bm r(e_3(\theta),\theta)$, and $\bm r(e_4(\theta),\theta)$ are five right eigenvectors corresponding to the five eigenvalues $1$, $e_1(\theta)$, $e_2(\theta)$, $e_3(\theta)$, and $e_4(\theta)$ of $\bm P_\theta$.  The derivation proceeds as before.

\subsection*{Proof of Lemma 1}

Assume either the S--S model or the W--S model.  Let $L_1,L_2,\ldots,L_n$ be i.i.d.~as $L$.  If we could show that $\P_{\theta}(L>x)$ is nondecreasing in $\theta$ for each $x\ge0$, then we would have
\begin{equation*}
\P_{\theta}(L_1+L_2+\cdots+L_n>x)
\end{equation*}
is nondecreasing in $\theta$ for each $x\ge0$, or equivalently $\P_\theta(\overline{L}>x)$ is nondecreasing in $\theta$ for each $x\ge0$.

It suffices to show that the tail probability \eqref{tail-eigen} is nondecreasing in $\theta$ for each positive integer $x\ge3$.  Alternatively, it is enough to show that
\begin{equation}\label{tail-matrixpower}
t(x,\theta):=\P_\theta(L\ge x)=1-[\bm P_\theta^{x-1}]_{\text{co},\text{7o}}
\end{equation}
is nondecreasing in $\theta$ for each positive integer $x\ge3$.

For small $x$, \eqref{tail-matrixpower} is simpler than \eqref{tail-eigen}.  For example,
\begin{equation*}
t(3,\theta)=\begin{cases}\dfrac{256+4\,\theta+\theta^2}{288}&(\text{S--S model}),\\
\noalign{\medskip}
\dfrac{288 + 27\,\theta - 2\,\theta^2}{324}&(\text{W--S model}),\end{cases}
\end{equation*}
which is clearly increasing in $\theta\in[0,1]$.

The formulation \eqref{tail-matrixpower} allows a computer-assisted proof for small and moderate $x$, say $x\le75$.  The point is that $t(x,\theta)$ is a polynomial in $\theta$ of degree at most $x-1$, hence its derivative with respect to $\theta$ is a polynomial in $\theta$ of degree at most $x-2$.  It suffice to show that this derivative is nonnegative.  Let us write it as
\begin{equation*}
\frac{\partial}{\partial\,\theta}t(x,\theta)=\sum_{i=0}^{x-2}a_{x,i}\,\theta^i.
\end{equation*}
Fix an integer $x\ge3$.  Suppose we can show $a_{x,0}>0$, $a_{x,1}>0$ and, for $i=2,3,\ldots,x-2$, the following hold:
\begin{enumerate}
\item \textit{One isolated negative coefficient}. If $a_{x,i-1}>0$, $a_{x,i}<0$, and $a_{x,i+1}\ge0$, then $a_{x,i-1}>|a_{x,i}|$ or $a_{x,i-2}>|a_{x,i}|$.

\item \textit{Two consecutive negative coefficients}. If $a_{x,i-2}>0$, $a_{x,i-1}<0$, $a_{x,i}<0$, and $a_{x,i+1}\ge0$, then $a_{x,i-2}>|a_{x,i-1}|+|a_{x,i}|$.

\item \textit{Three consecutive negative coefficients}. There are no such cases.
\end{enumerate}

In case 1, if $a_{x,i-1}>|a_{x,i}|$, we write
\begin{equation*}
a_{x,i-1}\,\theta^{i-1}+a_{x,i}\,\theta^i=(a_{x,i-1}+a_{x,i})\theta^{i-1}-a_{x,i}\,\theta^{i-1}(1-\theta),
\end{equation*}
which is nonnegative; if $a_{x,i-2}>|a_{x,i}|$, we write
\begin{align*}
&a_{x,i-2}\,\theta^{i-2}+a_{x,i-1}\,\theta^{i-1}+a_{x,i}\,\theta^i\\
&\quad{}=(a_{x,i-2}+a_{x,i})\theta^{i-2}+a_{x,i-1}\,\theta^{i-1}-a_{x,i}\,\theta^{i-2}(1-\theta^2),
\end{align*}
which is nonnegative.  In case 2, we write
\begin{align*}
&a_{x,i-2}\,\theta^{i-2}+a_{x,i-1}\,\theta^{i-1}+a_{x,i}\,\theta^i\\
&\quad{}=(a_{x,i-2}+a_{x,i-1}+a_{x,i})\theta^{i-2}-a_{x,i-1}\,\theta^{i-2}(1-\theta)-a_{x,i}\,\theta^{i-2}(1-\theta^2),
\end{align*}
which is nonnegative.  We conclude that $t(x,\theta)$ has a nonnegative derivative with respect to $\theta$ and is therefore nondecreasing in $\theta$.

Under the S--S model, we have verified these conditions on $x$ by computer for each $x\le75$.  Under the W--S model, we have done the same, but the condition $a_{x,1}>0$ fails for $x=3$, and condition 3 fails for $x=6,7,8,9$.  These five cases can easily be treated separately.
\medskip

For large $x$, we can use \eqref{tail-eigen} and the fact that the eight functions $e_1(\theta)$, $e_2(\theta)$, $e_3(\theta)$, $e_4(\theta)$, $c_1(\theta)$, $c_2(\theta)$, $c_3(\theta)$, and $c_4(\theta)$ are continuously differentiable to confirm by computer that
$(\partial/\partial\,\theta)t(x,\theta)\ge0$  for all $\theta\in[0,1]$ and for positive integers $x$ sufficiently large.  Factoring out $e_1(\theta)^{x-2}$, it suffices to show that
\begin{align}\label{tail-deriv}
&(x-1)c_1(\theta)e_1'(\theta)+c_1'(\theta)e_1(\theta)\nonumber\\
&\qquad{}+\sum_{i=2}^4[(x-1)c_i(\theta)e_i'(\theta)+c_i'(\theta)e_i(\theta)]\bigg(\frac{e_i(\theta)}{e_1(\theta)}\bigg)^{x-2}\ge0 
\end{align}
for all $\theta\in[0,1]$ and for positive integers $x$ sufficiently large.

For the S--S model, 
\begin{align*}
\min_\theta c_1(\theta)e_1'(\theta)&\ge0.0394,& \min_\theta c_1'(\theta)e_1(\theta)&\ge-0.0966, \\
\min_\theta c_2(\theta)e_2'(\theta)&\ge-0.0000464,& \min_\theta c_2'(\theta)e_2(\theta)&\ge0, \\
\min_\theta c_3(\theta)e_3'(\theta)&\ge-0.0000793,& \min_\theta c_3'(\theta)e_3(\theta)&\ge-0.00413, \\
\min_\theta c_4(\theta)e_4'(\theta)&\ge0.0142,& \min_\theta c_4'(\theta)e_4(\theta)&\ge0.00437, 
\end{align*}
and
\begin{equation*}
\max_\theta \frac{e_2(\theta)}{e_1(\theta)}\le0.860,\quad\max_\theta \frac{e_3(\theta)}{e_1(\theta)}\le0.823,\quad\min_\theta \frac{e_4(\theta)}{e_1(\theta)}\ge0.0754.
\end{equation*}
We conclude that \eqref{tail-deriv} holds for all positive integers $x\ge4$.

For the W--S model,
\begin{align*}
\min_\theta c_1(\theta)e_1'(\theta)&\ge0.0967,& \min_\theta c_1'(\theta)e_1(\theta)&\ge-0.141, \\
\min_\theta c_2(\theta)e_2'(\theta)&\ge-0.000575,& \min_\theta c_2'(\theta)e_2(\theta)&\ge0, \\
\min_\theta c_3(\theta)e_3'(\theta)&\ge-0.000489,& \min_\theta c_3'(\theta)e_3(\theta)&\ge-0.00297, \\
\min_\theta c_4(\theta)e_4'(\theta)&\ge-0.0104,& \min_\theta c_4'(\theta)e_4(\theta)&\ge0.0281, 
\end{align*}
and
\begin{equation*}
\max_\theta \frac{e_2(\theta)}{e_1(\theta)}\le0.873,\quad\max_\theta \frac{e_3(\theta)}{e_1(\theta)}\le0.843,\quad\max_\theta \frac{e_4(\theta)}{e_1(\theta)}\le0.252.
\end{equation*}
We conclude that \eqref{tail-deriv} holds for all positive integers $x\ge3$.  This completes the proof.

We acknowledge that this proof is not fully rigorous.  For example, the inequality $\min_\theta c_1(\theta)e_1'(\theta)\ge0.0394$ (S--S model) was confirmed as follows.  We used \textit{Mathematica} first to derive an explicit but complicated expression for $c_1(\theta)e_1'(\theta)$ and next to plot it as a function of $\theta\in[0,1]$.  The plot shows that the function is decreasing on $[0,1]$, and we can check that $c_1(0)e_1'(0)\approx0.0456155$ and $c_1(1)e_1'(1)\approx0.0394453$, resulting in the lower bound of 0.0394.  

The  weakness in this argument is the claimed monotonicity based on a visual inspection.

\end{document}